\documentclass[10pt]{article}
\usepackage{amsmath}
\usepackage{graphicx}
\usepackage{color}
\usepackage[table]{xcolor}
\usepackage{etoolbox} 
\usepackage{orcidlink}
\usepackage{hyperref}
\hypersetup{colorlinks=true, linkcolor=blue, citecolor=blue, urlcolor=blue}
\usepackage{caption}
\usepackage{subcaption}
\usepackage{setspace}
\usepackage{longtable}
\usepackage[export]{adjustbox}
\usepackage{authblk}
\usepackage{multirow}
\RequirePackage[numbers,sort&compress]{natbib}
\begin{document}

\begin{center}
\LARGE{Particle Dynamics and Thermal Properties in Kalb-Ramond ModMax Black Holes: Theoretical Predictions for Observational Tests of Exotic Physics}
\par\end{center}

\vspace{0.3cm}

\begin{center}
{\bf Ahmad Al-Badawi\orcidlink{0000-0002-3127-3453}}\footnote{\bf ahmadbadawi@ahu.edu.jo}\\
{\it Department of Physics, Al-Hussein Bin Talal University, 71111, Ma'an, Jordan}
\end{center}

\begin{center}
{\bf Faizuddin Ahmed\orcidlink{0000-0003-2196-9622}}\footnote{\bf faizuddinahmed15@gmail.com}\\
{\it Department of Physics, Royal Global University, Guwahati, 781035, Assam, India}
\end{center}

\begin{center}
{\bf \.{I}zzet Sakall{\i}\orcidlink{0000-0001-7827-9476}}\footnote{\bf izzet.sakalli@emu.edu.tr}\\
{\it Physics Department, Eastern Mediterranean University, Famagusta 99628, North Cyprus via Mersin 10, Turkey}
\end{center}

\vspace{0.3cm}

\begin{abstract}
We present a comprehensive theoretical study of geodesic motion and thermodynamic behavior in Kalb--Ramond (KR) black hole (BH) spacetimes sourced by ModMax electrodynamics. Both neutral and charged test particle dynamics are investigated, highlighting how the Lorentz symmetry breaking (LSB) parameter $\ell$, the ModMax nonlinearity parameter $\gamma$, and the discrete branch parameter $\zeta$ significantly modify orbital structures compared to classical Schwarzschild and Reissner--Nordstr\"{o}m (RN) solutions. Effective potential analysis reveals notable shifts in the innermost stable circular orbit (ISCO): ordinary branches allow stable orbits closer to the horizon, while phantom branches shift them outward by factors of 5--10. For charged particles, the combined influence of modified gravity and nonlinear electromagnetic fields may induce chaotic trajectories in certain regimes. On the thermodynamic side, we derive full expressions for Hawking temperature, entropy, and Helmholtz free energy. Ordinary branches exhibit divergent specific heat indicating second-order phase transitions, whereas phantom branches yield consistently negative specific heat, implying thermal instability. Phantom BHs are found to possess higher Hawking temperatures and show distinct thermodynamic phase structures, including Hawking--Page-type transitions. Observational features such as BH shadows and gravitational lensing are explored, revealing parameter-dependent changes in photon sphere radii and deflection angles. Notably, the deflection angle analysis via the Gauss--Bonnet theorem method (GBTm) shows opposite-sign electromagnetic corrections for phantom versus ordinary branches, suggesting potential observational discriminants.

\end{abstract}

% \keywords{Kalb-Ramond gravity; ModMax Electrodynamics; Geodesic structure; Thermodynamics; deflection angle.}

\section{Introduction} \label{isec1}
The exploration of exotic BH solutions in modified theories of gravity has emerged as one of the most vibrant and theoretically rich areas of contemporary gravitational physics, offering unprecedented opportunities to probe the fundamental nature of spacetime, symmetry principles, and electromagnetic interactions in extreme gravitational environments \cite{isz01,isz02,isz03}. Among the diverse landscape of alternative gravitational theories, those incorporating LSB effects and nonlinear electrodynamics have garnered particular attention due to their potential to address longstanding puzzles in both fundamental physics and observational cosmology. The theoretical framework combining KR gravity with ModMax electrodynamics represents a particularly fascinating confluence of these exotic physics concepts, creating BH solutions that exhibit rich phenomenological behavior absent in classical General Relativity (GR) while maintaining mathematical elegance and physical consistency \cite{isz04,isz05,isz06}.

LSB, arising from the spontaneous breaking of Lorentz invariance at high energy scales, has emerged as a compelling theoretical possibility that could resolve several outstanding problems in particle physics and cosmology, including the hierarchy problem, the cosmological constant puzzle, and the nature of dark energy \cite{isz07,isz08,isz09,isz09x1,isz09x2,isz09x3,isz09x4,isz09x5,isz09x6,isz09x7}. The KR field, originally introduced in string theory as an antisymmetric tensor field, provides a natural mechanism for LSB when it acquires a non-zero vacuum expectation value, creating preferred reference frames and modifying the causal structure of spacetime in ways that could have profound observational consequences \cite{isz10,isz11,isz12}. The incorporation of KR fields into gravitational theories leads to modified Einstein field equations that support novel BH solutions with altered horizon structures, thermodynamic properties, and particle dynamics compared to their General Relativistic counterparts.

Simultaneously, the study of nonlinear electrodynamics has experienced a renaissance driven by both theoretical motivations from quantum field theory and observational hints of electromagnetic phenomena that transcend the linear Maxwell regime \cite{isz13,isz14,isz15}. ModMax electrodynamics, proposed as a conformal and duality-invariant extension of Maxwell theory, represents one of the most elegant formulations of nonlinear electrodynamics, preserving essential symmetries while introducing controlled deviations from linearity that could manifest in strong electromagnetic field environments such as those found near charged BHs \cite{isz16,isz17,isz18}. The ModMax theory exhibits remarkable properties including electromagnetic duality invariance and conformal symmetry, making it an attractive candidate for describing electromagnetic phenomena in the strong-field regime where conventional Maxwell theory might break down.

The marriage of KR gravity and ModMax electrodynamics creates a theoretical framework of exceptional richness, where LSB effects and electromagnetic nonlinearity combine to produce BH solutions that challenge our understanding of spacetime structure and thermodynamics \cite{isz19,isz20,isz21}. Recent investigations have revealed that these KR ModMax BHs exhibit parameter-dependent modifications to all aspects of BH physics, from the basic metric structure and horizon formation to the complex dynamics of particle motion and thermal radiation \cite{isz47}. These exotic solutions interpolate between classical Schwarzschild and RN geometries while introducing genuinely novel physics through the interplay of LSB violations and electromagnetic nonlinearity. The theory introduces several fundamental parameters—the LSB parameter $\ell$, the ModMax nonlinearity parameter $\gamma$, and the discrete branch parameter $\zeta$—that control the strength of exotic effects and create a rich parameter space for exploring deviations from classical predictions.

The geodesic structure analysis of test particle motion in KR ModMax BH spacetimes reveals profound modifications to orbital dynamics that extend far beyond simple parameter rescaling effects found in many alternative theories \cite{isz22,isz23,isz24,isz24x1,isz24x2,isz24x3,isz24x4}. Neutral particles, despite being electromagnetically inactive, experience altered gravitational interactions due to the modified spacetime geometry arising from LSB effects and the gravitational backreaction of nonlinear electromagnetic fields. The effective potentials governing particle motion exhibit complex parameter-dependent structures that can dramatically shift the locations of stable circular orbits, modify the innermost stable circular orbit
 (ISCO) radius, and create novel trajectory patterns that have no analogue in GR. For charged particles, the situation becomes even more intricate, as they respond to both the modified gravitational field and the nonlinear electromagnetic environment, potentially leading to chaotic dynamics and complex phase space structures that could have profound implications for accretion processes and high-energy particle acceleration mechanisms \cite{isz25,isz26,isz27}.

The thermodynamic properties of KR ModMax BHs provide another fascinating arena where exotic physics manifests through modifications to fundamental quantities such as Hawking temperature, entropy, and specific heat capacity \cite{isz28,isz29,isz30}. The interplay between LSB effects and ModMax electrodynamics creates branch-dependent thermal behaviors, where ordinary and phantom configurations exhibit qualitatively different thermodynamic signatures including altered stability properties, modified phase transition structures, and distinctive temperature-mass relationships. These thermodynamic modifications not only provide theoretical insights into the statistical mechanical foundations of BH physics in modified gravity theories but also offer potential observational targets for distinguishing exotic BH solutions from their classical counterparts through precision measurements of thermal radiation spectra and correlation functions.

The study of BH shadows and gravitational lensing in the KR ModMax framework opens additional observational windows for testing exotic physics through precision astronomical measurements \cite{isz31,isz32,isz33}. Recent breakthrough observations by the Event Horizon Telescope have demonstrated the feasibility of direct BH shadow measurements with unprecedented precision, creating opportunities to constrain modified gravity theories through comparison of predicted and observed shadow characteristics. The KR ModMax theory predicts distinctive parameter-dependent modifications to shadow sizes and shapes that could serve as smoking gun signatures of LSB violations and electromagnetic nonlinearity, particularly when combined with systematic studies across multiple BH systems with varying masses and charges.

Gravitational lensing provides another powerful probe of spacetime geometry modifications, where the deflection of light by KR ModMax BHs encodes information about both LSB effects and electromagnetic nonlinearity through higher-order corrections to the classical deflection angle \cite{isz34,isz35,isz36}. The systematic analysis of lensing signatures across different impact parameters and observing geometries could potentially provide tight constraints on fundamental theory parameters while testing the consistency of exotic physics across different energy scales and field strengths. The branch-dependent nature of electromagnetic contributions creates particularly distinctive signatures, where ordinary and phantom configurations produce opposite-sign corrections that could be unambiguously distinguished through precision astrometric measurements.

Our primary motivation for this comprehensive investigation stems from the recognition that the confluence of LSB effects and nonlinear electrodynamics in the KR ModMax framework creates a uniquely rich theoretical laboratory for exploring fundamental physics while maintaining direct connections to potential observational signatures \cite{isz37,isz38,isz39}. The theory's mathematical elegance, combined with its prediction of distinctive phenomenological effects across multiple observational channels, makes it an ideal candidate for systematic theoretical development and observational constraint. Furthermore, the parameter space of the theory is sufficiently rich to accommodate a wide range of exotic effects while remaining tractable for detailed analytical and numerical investigation. The central aims of this work are manifold and interconnected. First, we seek to provide a comprehensive theoretical framework for understanding particle dynamics in KR ModMax BH spacetimes, encompassing both neutral and charged test particles across the full parameter space of the theory. This includes detailed analysis of effective potentials, ISCO locations, orbital stability, and trajectory patterns, with particular emphasis on identifying distinctive signatures that distinguish this exotic framework from classical GR and other alternative theories. Second, we aim to establish a complete thermodynamic description of KR ModMax BHs, including the derivation and analysis of fundamental quantities such as mass, temperature, entropy, and specific heat, with careful attention to the branch-dependent behaviors that arise from the discrete parameter $\zeta$ \cite{isz40}.

Third, we investigate the observational signatures of KR ModMax BHs through detailed analysis of shadow properties and gravitational lensing effects, providing theoretical predictions that can be directly compared with current and future astronomical observations. This includes systematic exploration of parameter dependencies and identification of optimal observational strategies for constraining or detecting exotic physics effects. Fourth, we seek to establish the theoretical foundations for future phenomenological studies by developing analytical approximations, numerical methods, and scaling relationships that can be applied to realistic astrophysical scenarios involving KR ModMax BHs in various environments and evolutionary stages.

The paper is organized as follows: In Section \ref{isec2}, we review the fundamental properties of KR ModMax BH solutions, including their metric structure, horizon analysis, and limiting cases that connect to well-known classical results. Section \ref{isec3} presents a comprehensive analysis of neutral test particle geodesics, covering effective potentials, circular orbits, ISCO determination, and trajectory patterns across the theory's parameter space. Moreover, we extend the geodesic analysis to charged particles, exploring the complex interplay between gravitational and electromagnetic effects and their implications for orbital dynamics and stability. Section \ref{isec5} provides a detailed investigation of the thermodynamic properties of KR ModMax BHs, including temperature, entropy, specific heat, and phase transition analysis for both ordinary and phantom branch configurations. Section \ref{isec6} examines the shadow properties of these exotic BHs, deriving photon sphere locations and shadow radii while comparing results across different theoretical limits. Section \ref{isec7} presents the gravitational lensing analysis using the GBTm, computing deflection angles and exploring their parameter dependencies and observational implications. Finally, Section \ref{isec8} summarizes our main findings, discusses their broader implications for fundamental physics and observational astronomy, and outlines promising directions for future theoretical and observational investigations.

\section{Review of KR ModMax BH} \label{isec2}

The theoretical framework underlying this investigation centers on a recently discovered class of charged BH solutions that emerge from the intricate interplay between LSB effects and nonlinear electromagnetic phenomena. These solutions, first reported in Ref. \cite{sec2is01}, represent a significant advancement in our understanding of modified gravity theories where ModMax electrodynamics is non-minimally coupled to a background KR two-form field. The resulting spacetime geometry exhibits rich phenomenology that deviates substantially from classical GR predictions, offering new avenues for testing fundamental physics in strong gravitational regimes.

The line element characterizing this exotic spacetime can be expressed in the standard spherically symmetric form:
\begin{equation}
ds^2=-F(r)\,dt^2+\frac{dr^2}{F(r)}+r^2 d\theta^2+r^2 \sin^2\theta d\phi^2,
\label{aa1}
\end{equation}
where the metric function $F(r)$ encapsulates the novel physics arising from the coupled KR-ModMax system:
\begin{equation}
F(r)=\frac{1}{1-\ell}-\frac{2M}{r}+\zeta\frac{e^{-\gamma }  \,Q^2}{(1-\ell)^2 r^2}.
\label{aa2}
\end{equation}

The parameter $\ell$ serves as a dimensionless measure of LSB strength, originating from the non-zero vacuum expectation value of the KR field that permeates spacetime \cite{isz04,isz05}. This background field configuration fundamentally alters the causal structure of spacetime, leading to preferred reference frames and modified dispersion relations for propagating fields. The discrete parameter $\zeta$ distinguishes between ordinary ($\zeta=1$) and phantom ($\zeta=-1$) branches, each exhibiting distinct physical characteristics and stability properties \cite{isz06}. The ModMax parameter $\gamma$ quantifies the degree of nonlinearity in the electromagnetic sector, while $Q$ represents the electric charge of the BH \cite{isz17,sec2is06}.

The horizon structure of this exotic spacetime reveals fascinating complexity, with the possible horizon radii given by:
\begin{equation}
r_\pm=M\,(1-\ell)\left[1\pm \sqrt{1-\frac{\zeta\,e^{-\gamma}\,Q^2}{M^2\,(1-\ell)^3}}\right]
\label{Horizons_RN}
\end{equation}

Several illuminating special cases emerge from this general framework. When $Q = 0$, the solution reduces to a pure KR gravity BH with a single horizon at $r = 2M(1 - \ell)$, demonstrating how LSB modifies even neutral spacetimes. For $\ell = 0$ and $\zeta = 1$, we recover ModMax electrodynamics without LSB, yielding horizons at $r_\pm = M \left[1 \pm \sqrt{1 - \frac{e^{-\gamma} Q^2}{M^2}} \right]$. The case $\gamma = 0$ and $\zeta = 1$ corresponds to an Reissner-Nordstr\"{o}m (RN) BH in KR gravity, with horizons located at $r_\pm = M(1 - \ell) \left[1 \pm \sqrt{1 - \frac{Q^2}{M^2(1 - \ell)^3}} \right]$.

In the extremal limit where inner and outer horizons coincide, the degenerate horizon radius satisfies:
\begin{equation}
M=\frac{e^{-\gamma}\,Q^2}{(1-\ell)^2\,r_\text{ext}}\quad,\quad r_\text{ext}=\pm\,e^{-\gamma/2}\,\left|\frac{Q}{1-\ell}\right|.
\label{ext}
\end{equation}

\begin{figure}[ht!]  
\includegraphics[width=0.4\linewidth]{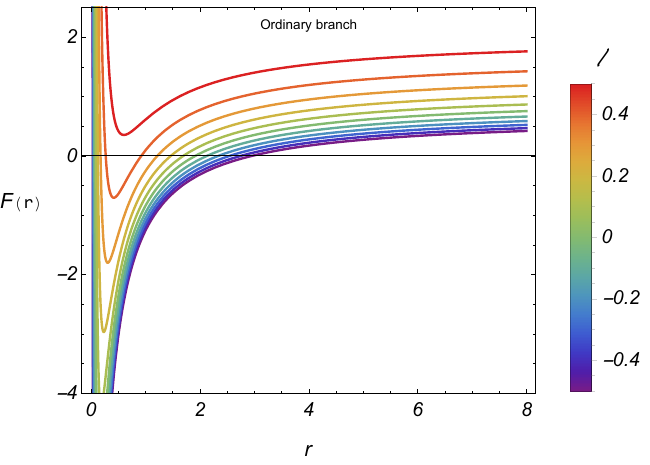}\quad\quad\quad
\includegraphics[width=0.4\linewidth]{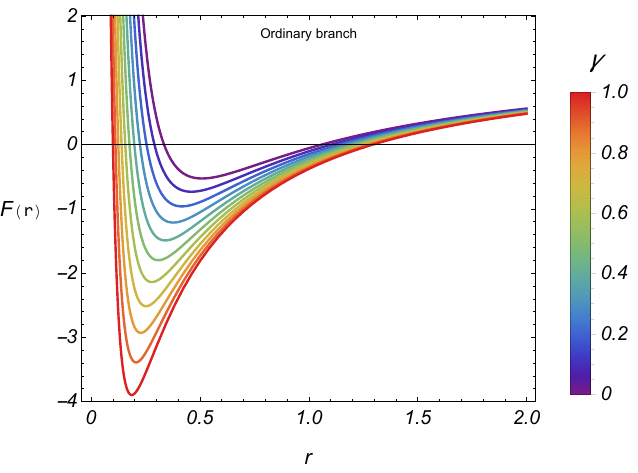}
\caption{\footnotesize Plot of the metric function $F(r)$ versus the parameters $\ell$ (left) and $\gamma$ (right). Here, $M=1$ and $q=0.5$.}
\label{figa1}
\end{figure}

\begin{figure}[ht!]    
\includegraphics[width=0.4\linewidth]{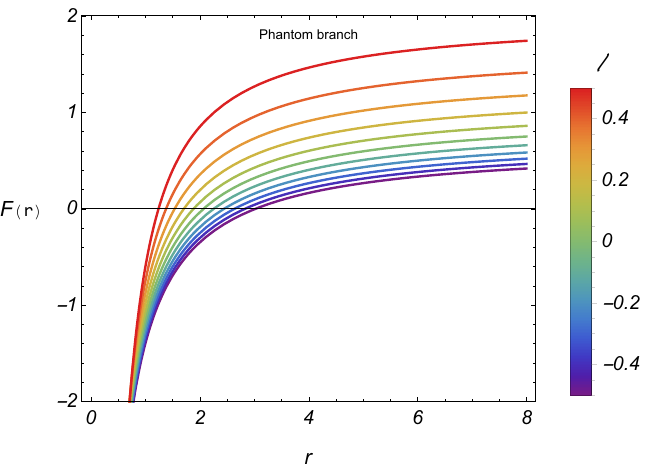}\quad\quad\quad
\includegraphics[width=0.4\linewidth]{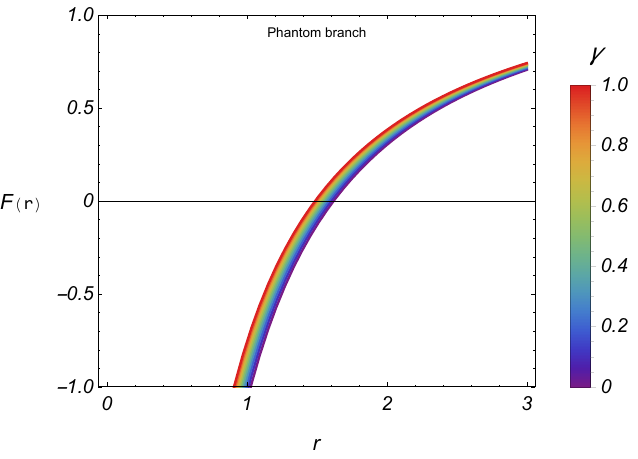}
\caption{\footnotesize Plot of the metric function $F(r)$ versus the parameters $\ell$ (left) and $\gamma$ (right). Here, $M=1$ and $q=0.5$.}
\label{figa2}
\end{figure}

Figures \ref{figa1} and \ref{figa2} provide comprehensive visualization of how the metric function $F(r)$ responds to variations in the fundamental parameters $\ell$ and $\gamma$ for both ordinary and phantom BH branches. The ordinary branch (Figure \ref{figa1}) exhibits a characteristic behavior where increasing $\ell$ shifts the metric function upward, effectively reducing the gravitational strength near the BH, while the ModMax parameter $\gamma$ introduces subtle but significant modifications to the electromagnetic contributions. In contrast, the phantom branch (Figure \ref{figa2}) displays markedly different characteristics, with the LSB parameter $\ell$ producing more dramatic alterations to the spacetime geometry. The phantom branch typically supports only a single horizon, whereas the ordinary branch can accommodate up to two horizons depending on the parameter values. These plots clearly demonstrate the rich parameter space available in this theory and highlight the fundamental differences between ordinary and phantom configurations, which have profound implications for BH formation, evolution, and observational signatures.

\begin{table}[ht!]
\centering
\renewcommand{\arraystretch}{1.2}
\setlength{\tabcolsep}{6pt}
\begin{tabular}{|c|c|c|c|}
\hline
\textbf{$\gamma$} & \textbf{$\ell$} & \textbf{$\zeta$} & \textbf{Horizon(s)} \\
\hline
\multirow{3}{*}{0} 
& 0      & 0   & $[0.0,\ 2.0]$ \\
&        & 1   & $[1.0,\ 1.0]$ \\
&        & -1  & $[2.414213562]$ \\
\cline{2-4}
& 0.01   & 0   & $[0.0,\ 1.980]$ \\
&        & 1   & $[\,]$ \\
&        & -1  & $[2.400744843]$ \\
\cline{2-4}
& 0.1    & 0   & $[0.0,\ 1.800]$ \\
&        & 1   & $[\,]$ \\
&        & -1  & $[2.286041526]$ \\
\hline
\multirow{4}{*}{0.5} 
& 0      & 0   & $[0.0,\ 2.0]$ \\
&        & 1   & $[0.3727,\ 1.6273]$ \\
&        & -1  & $[2.267489905]$ \\
\cline{2-4}
& 0.01   & 0   & $[0.0,\ 1.980]$ \\
&        & 1   & $[0.3838,\ 1.5962]$ \\
\cline{2-4}
& 0.1    & 0   & $[0.0,\ 1.800]$ \\
&        & 1   & $[0.5311,\ 1.2689]$ \\
&        & -1  & $[2.118163764]$ \\
\hline
\multirow{4}{*}{1} 
& 0      & 0   & $[0.0,\ 2.0]$ \\
&        & 1   & $[0.2049,\ 1.7951]$ \\
&        & -1  & $[2.169563782]$ \\
\cline{2-4}
& 0.01   & 0   & $[0.0,\ 1.980]$ \\
&        & 1   & $[0.2099,\ 1.7701]$ \\
&        & -1  & $[2.152624357]$ \\
\cline{2-4}
& 0.1    & 0   & $[0.0,\ 1.800]$ \\
&        & 1   & $[0.2666,\ 1.5334]$ \\
&        & -1  & $[2.003972343]$ \\
\hline
\end{tabular}
\caption{\footnotesize Parameter-dependent horizon structure in KR ModMax BH spacetimes. For each combination of ModMax parameter $\gamma$, LSB parameter $\ell$, and branch selector $\zeta$, the corresponding horizon radii are listed. The systematic variation reveals how exotic physics modifications alter the causal structure compared to classical BH solutions. Parameters: $M=1$ and $Q=1$.}
\label{gltab}
\end{table}

The comprehensive horizon analysis presented in Table \ref{gltab} reveals the intricate relationship between the theory's fundamental parameters and the resulting BH horizon structure. The table systematically explores various combinations of the ModMax parameter $\gamma$, LSB parameter $\ell$, and branch selector $\zeta$, demonstrating the rich phenomenology accessible within this theoretical framework. Notably, certain parameter combinations result in the absence of horizons (indicated by empty brackets), suggesting the formation of naked singularities or other exotic spacetime configurations. The phantom branch ($\zeta = -1$) consistently produces single horizons across all tested parameter values, reflecting its unique gravitational characteristics. In contrast, the ordinary branch ($\zeta = 1$) exhibits a complex pattern where some configurations support dual horizons while others completely lack horizon formation. The systematic variation in horizon locations with changing parameters provides crucial insights for understanding the stability and physical viability of these exotic BH solutions, with direct implications for their potential observational signatures and astrophysical relevance.

\begin{figure*}[ht!]
    \centering
    \setlength{\tabcolsep}{0pt}

    % Row 1
    \begin{minipage}{0.14\textwidth}
         \centering
             \includegraphics[width=\textwidth]{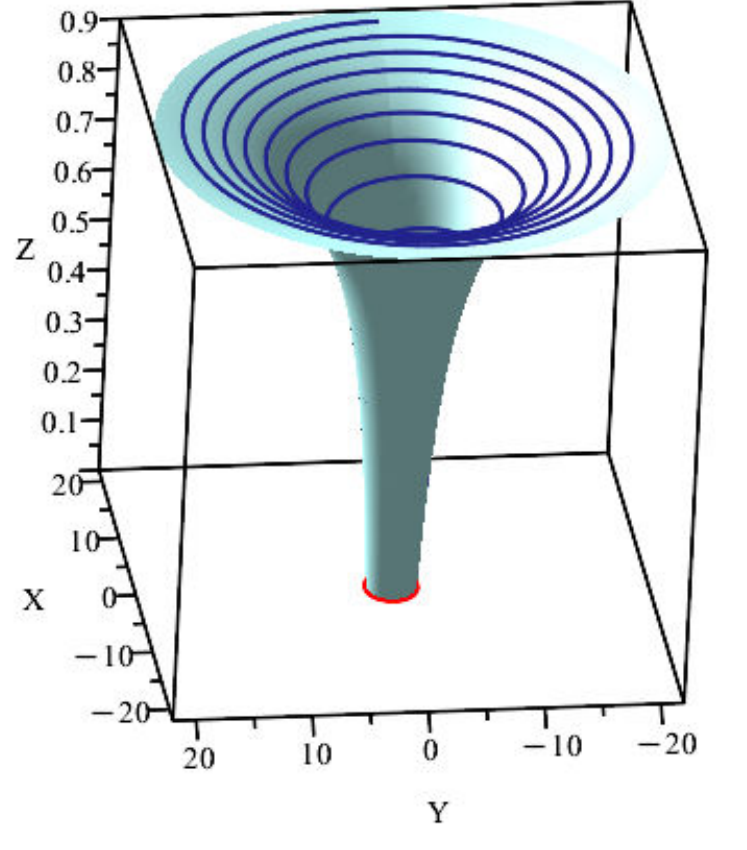}
        \subcaption{\parbox[t]{3.5cm}{[$\ell=0$, $\zeta=0$, $\gamma=0$]}}
        \label{fig:fig1}
    \end{minipage}
     \hspace{5em}
    \begin{minipage}{0.14\textwidth}
        \centering
           \includegraphics[width=\textwidth]{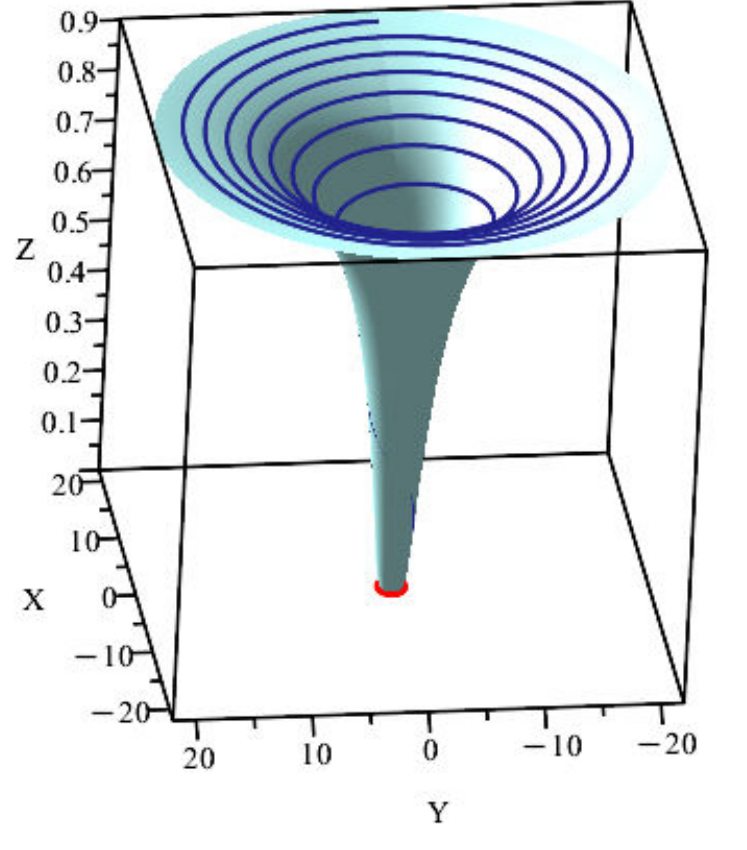}
        \subcaption{\parbox[t]{3.5cm}{[$\ell=0$, $\zeta=1$, $\gamma=0$]}}
        \label{fig:fig2}
    \end{minipage}
    \hspace{5em}
    \begin{minipage}{0.14\textwidth}
       \centering
    \includegraphics[width=\textwidth]{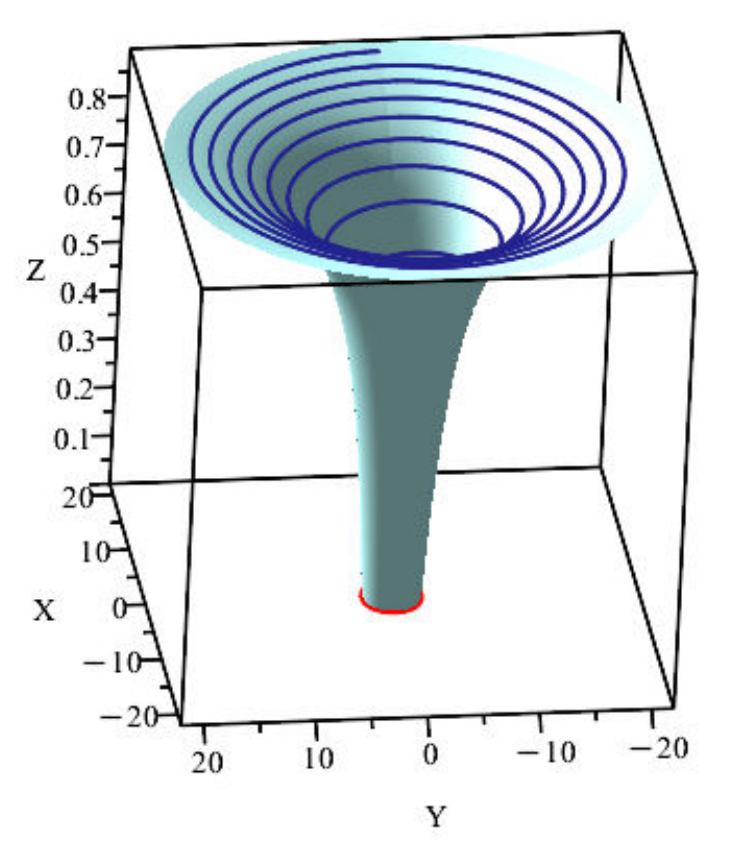}
        \subcaption{\parbox[t]{3.5cm}{[$\ell=0$, $\zeta=-1$, $\gamma=0$]}}
        \label{fig:fig3}
    \end{minipage}

    \vspace{0.6em}

    % Row 2
    \begin{minipage}{0.14\textwidth}
        \centering
        \includegraphics[width=\textwidth]{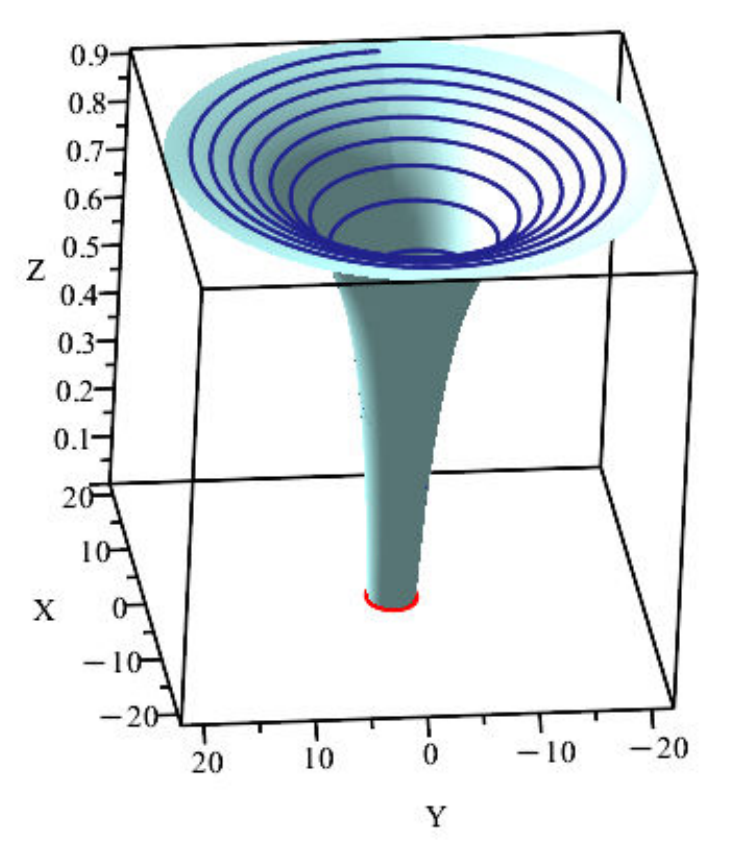}
        \subcaption{\parbox[t]{3.5cm}{[$\ell=0.01$, $\zeta=0$, $\gamma=0$]}}
        \label{fig:fig4}
    \end{minipage}
  \hspace{5em}
    \begin{minipage}{0.14\textwidth}
        \centering
        \includegraphics[width=\textwidth]{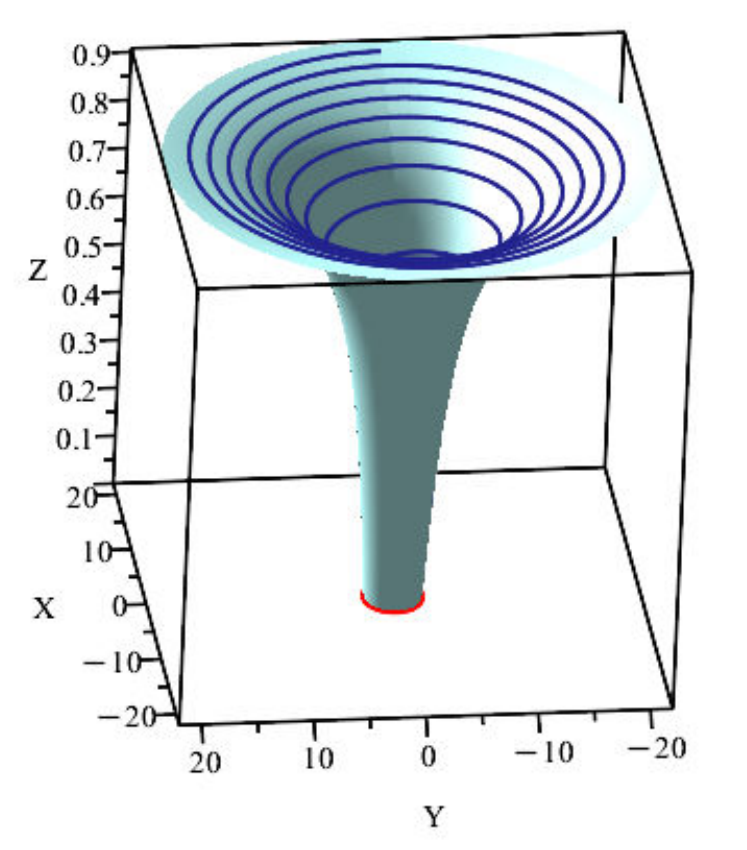}
        \subcaption{\parbox[t]{3.5cm}{[$\ell=0.01$, $\zeta=-1$, $\gamma=0$]}}
        \label{fig:fig5}
    \end{minipage}
   \hspace{5em}
    \begin{minipage}{0.14\textwidth}
        \centering
        \includegraphics[width=\textwidth]{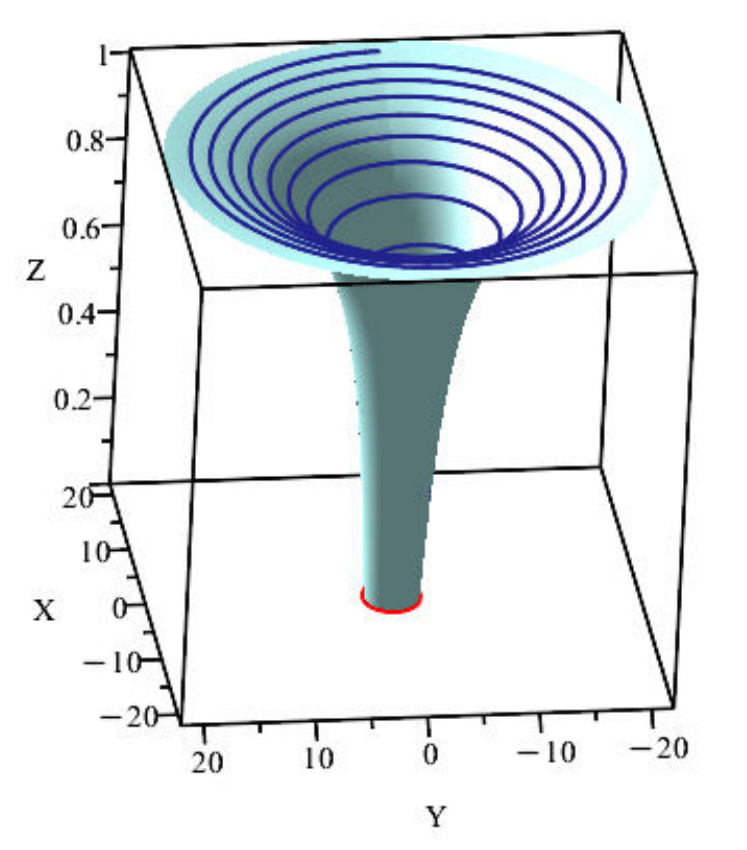}
        \subcaption{\parbox[t]{3.5cm}{[$\ell=0.1$, $\zeta=-1$, $\gamma=0$]}}
        \label{fig:fig6}
    \end{minipage}

    \vspace{0.6em}

    % Row 3
    \begin{minipage}{0.14\textwidth}
        \centering
        \includegraphics[width=\textwidth]{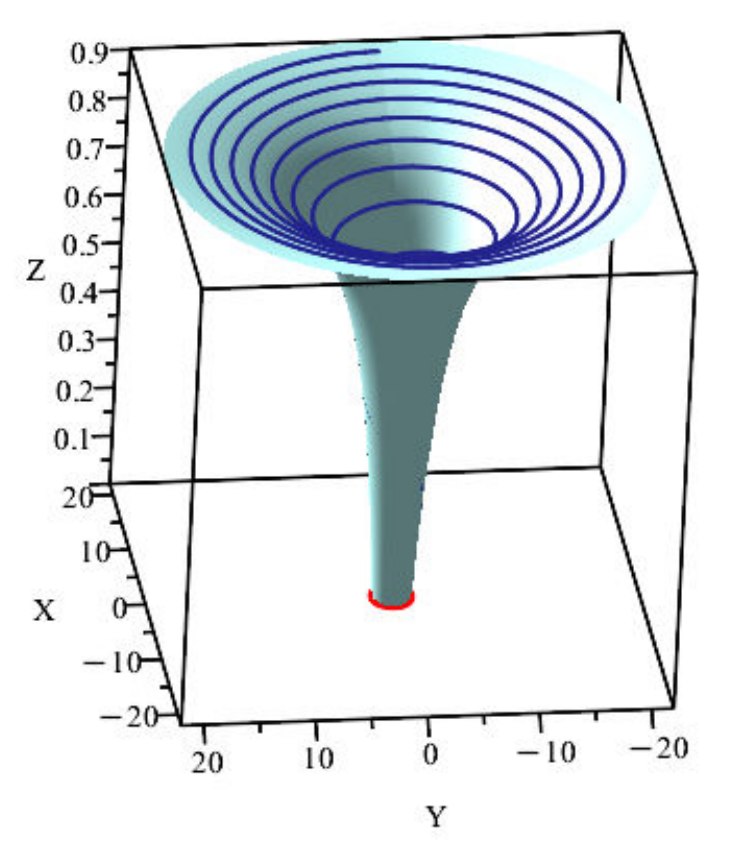}
        \subcaption{\parbox[t]{3.5cm}{[$\ell=0$, $\zeta=1$, $\gamma=0.5$]}}
        \label{fig:fig7}
    \end{minipage}
    \hspace{5em}
    \begin{minipage}{0.14\textwidth}
        \centering
        \includegraphics[width=\textwidth]{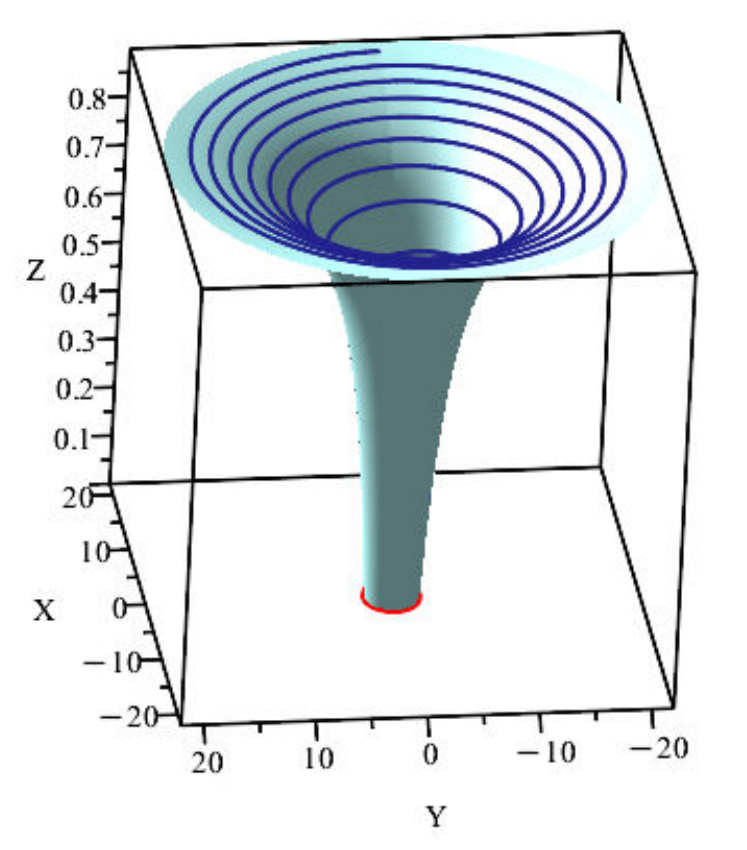}
        \subcaption{\parbox[t]{3.5cm}{[$\ell=0$, $\zeta=-1$, $\gamma=0.5$]}}
        \label{fig:fig8}
    \end{minipage}
    \hspace{5em}
    \begin{minipage}{0.14\textwidth}
        \centering
        \includegraphics[width=\textwidth]{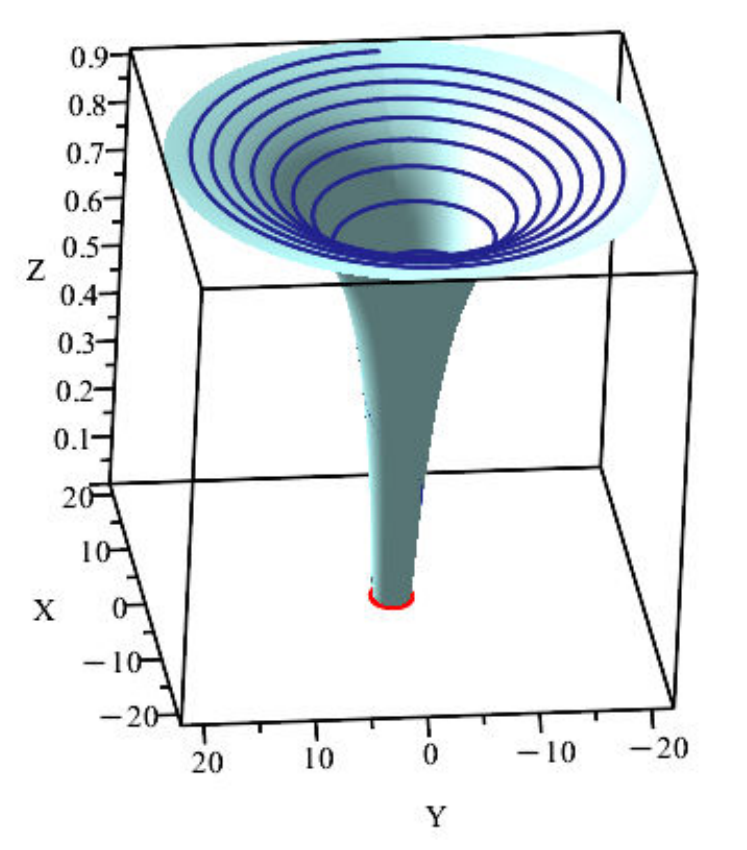}
        \subcaption{\parbox[t]{3.5cm}{[$\ell=0.01$, $\zeta=1$, $\gamma=0.5$]}}
        \label{fig:fig9}
    \end{minipage}

    \vspace{0.6em}

    % Row 4
    \begin{minipage}{0.15\textwidth}
        \centering
        \includegraphics[width=\textwidth]{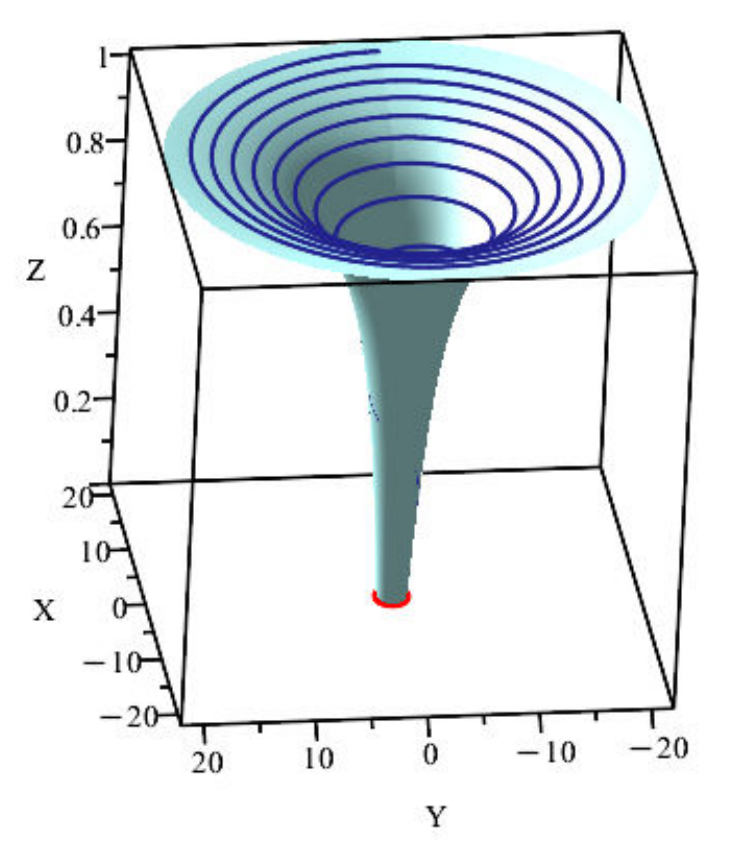}
       \subcaption{\parbox[t]{3.5cm}{[$\ell=0.1$, $\zeta=1$, $\gamma=0.5$]}}
        \label{fig:fig10}
    \end{minipage}
    \hspace{5em}
    \begin{minipage}{0.15\textwidth}
        \centering
        \includegraphics[width=\textwidth]{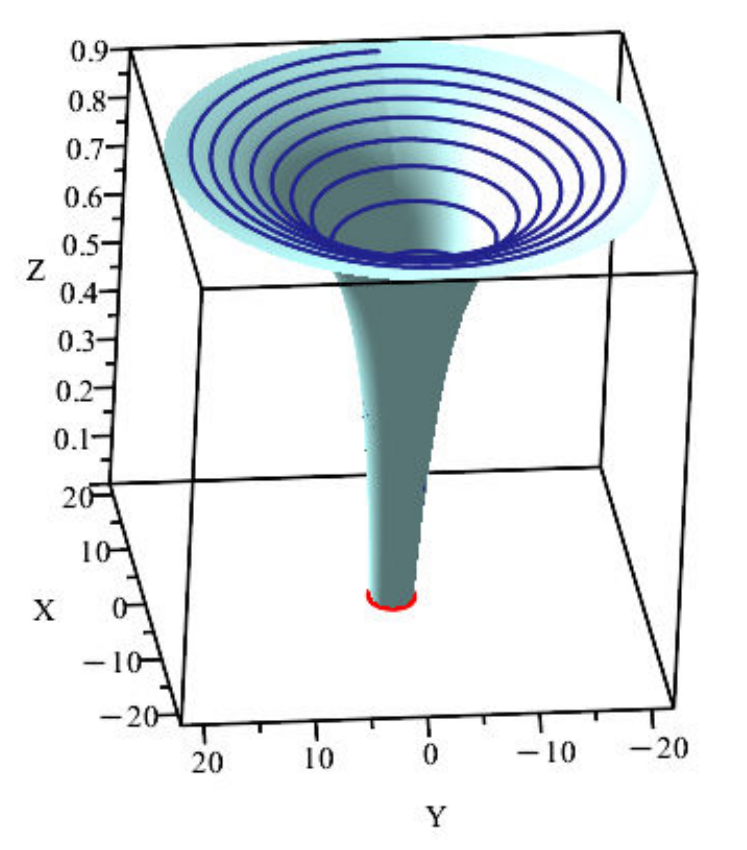}
        \subcaption{\parbox[t]{3.5cm}{[$\ell=0$, $\zeta=1$, $\gamma=1$]}}
        \label{fig:fig11}
    \end{minipage}
    \hspace{5em}
    \begin{minipage}{0.15\textwidth}
        \centering
        \includegraphics[width=\textwidth]{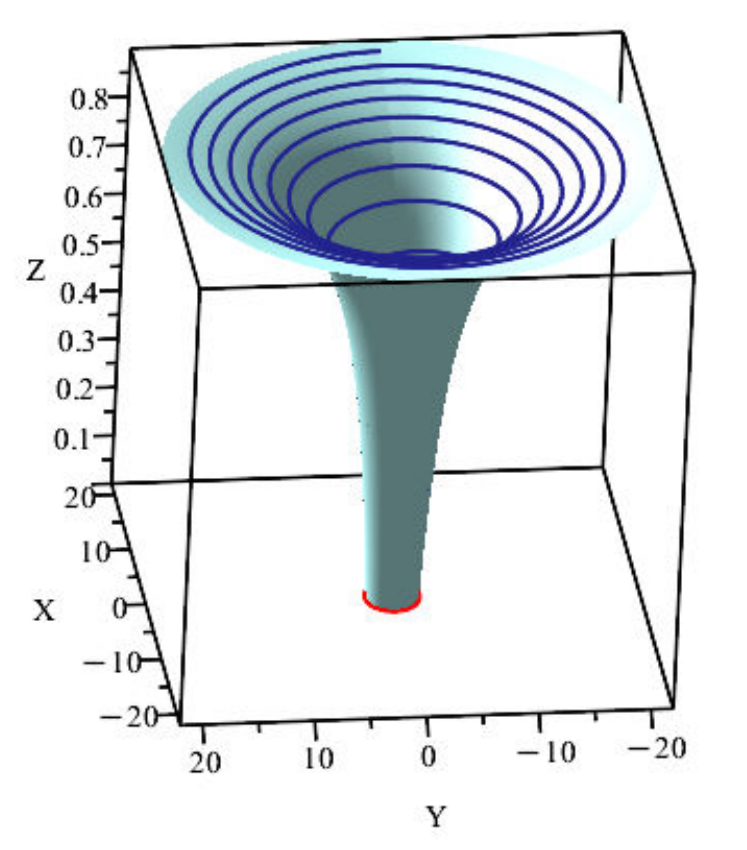}
        \subcaption{\parbox[t]{3.5cm}{[$\ell=0$, $\zeta=-1$, $\gamma=1$]}}
        \label{fig:fig12}
    \end{minipage}

    \vspace{0.6em}

    % Row 5
    \begin{minipage}{0.15\textwidth}
        \centering
        \includegraphics[width=\textwidth]{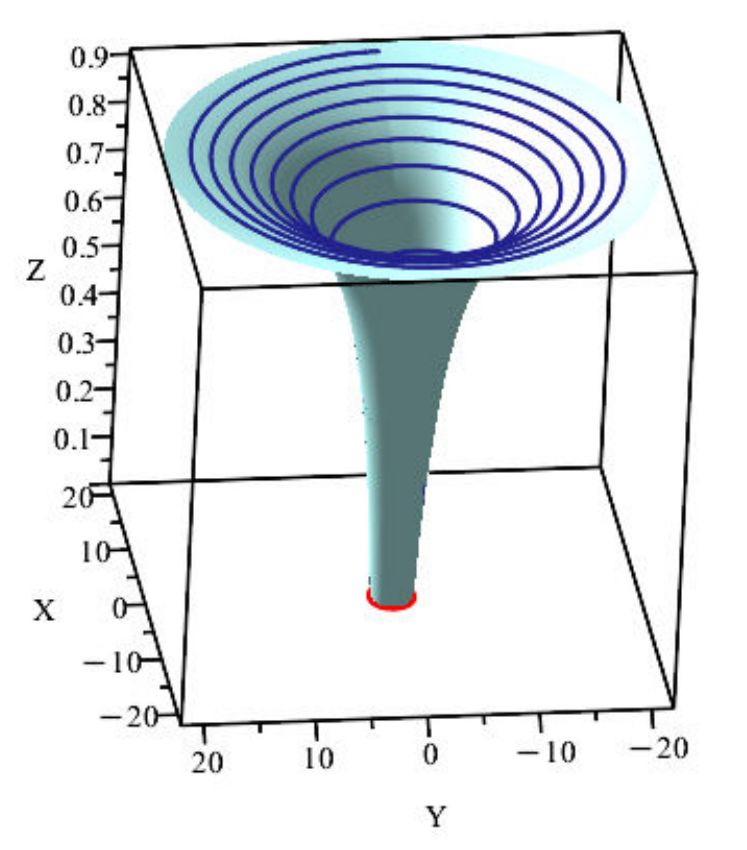}
       \subcaption{\parbox[t]{3.5cm}{[$\ell=0.01$, $\zeta=1$, $\gamma=1$]}}
        \label{fig:fig13}
    \end{minipage}
    \hspace{5em}
    \begin{minipage}{0.15\textwidth}
        \centering
        \includegraphics[width=\textwidth]{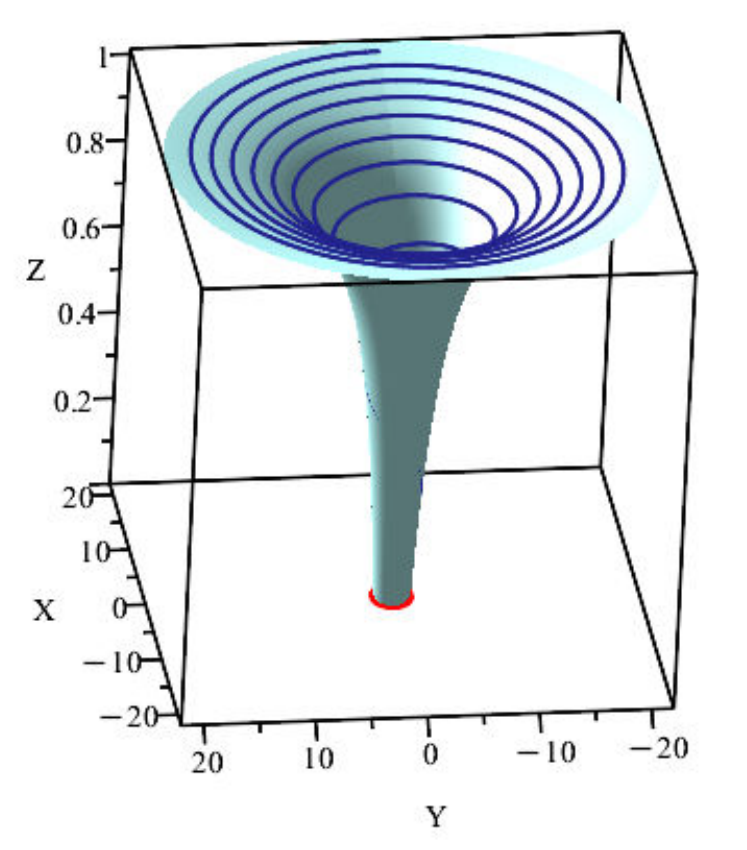}
        \subcaption{\parbox[t]{3.5cm}{[$\ell=0.1$, $\zeta=1$, $\gamma=1$]}}
        \label{fig:fig14}
    \end{minipage}
    \hspace{5em}
    \begin{minipage}{0.15\textwidth}
        \centering
        \includegraphics[width=\textwidth]{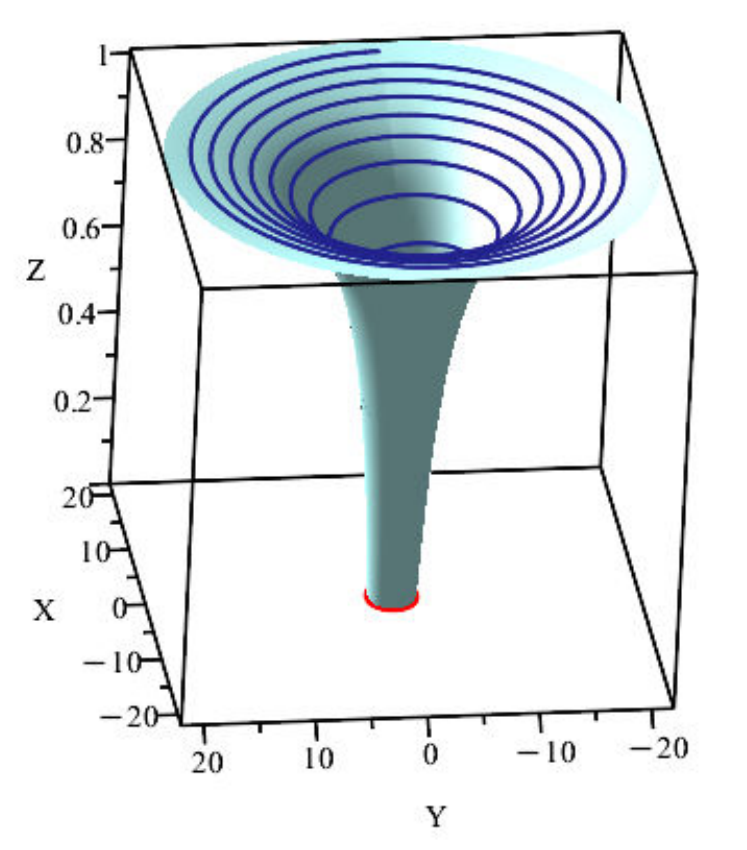}
        \subcaption{\parbox[t]{3.5cm}{[$\ell=1$, $\zeta=-1$, $\gamma=1$]}}
        \label{fig:fig15}
    \end{minipage}

    \caption{\footnotesize Embedding diagrams of the KR ModMax BH for various parameter values of $\ell$, $\zeta$, and $\gamma$. The mass and coupling parameters are set to $M=1$ and $Q=1$. The horizons (red rings) are governed by the horizons served in Table \ref{gltab}.}
    \label{fig:isfull_embedding}
\end{figure*}

The embedding diagrams presented in Figure \ref{fig:isfull_embedding} provide an intuitive three-dimensional visualization of how the KR ModMax BH geometry evolves across the theory's parameter space. These diagrams represent spatial slices of the spacetime, revealing the characteristic funnel-like structure that defines BH geometries, with the red rings marking the horizon locations as determined by Table \ref{gltab}. The systematic progression through different parameter combinations demonstrates the remarkable flexibility of this theoretical framework in producing diverse spacetime configurations. The top row showcases baseline configurations with minimal parameter variations, establishing reference geometries for comparison. As we progress through increasing values of $\ell$ and $\gamma$, the throat structures exhibit notable modifications in their depth, width, and overall shape, reflecting the profound influence of LSB and ModMax effects on spacetime curvature. The phantom branch configurations display particularly striking deviations from standard BH geometries, often featuring more pronounced asymmetries and altered causal structures. These visual representations not only enhance our understanding of the underlying physics but also provide valuable insights for potential observational strategies, as the geometric modifications would directly impact phenomena such as gravitational lensing, particle orbits, and electromagnetic signatures around these exotic BHs \cite{sec2is07,isz31}.

\section{Geodesic Motion} \label{isec3}

\subsection{Neutral Test Particles}

The study of geodesic motion in the KR ModMax BH spacetime provides fundamental insights into how LSB effects and nonlinear electrodynamics influence particle dynamics in strong gravitational fields. In the spacetime of a charged BH, neutral particles follow trajectories determined solely by spacetime curvature, as they remain unaffected by electromagnetic interactions \cite{isz22,isz23}. However, the presence of both the electric charge and the ModMax parameter, combined with LSB effects, introduces additional geometric structure that significantly alters the effective potential governing particle motion. This modification results in neutral particles experiencing gravitational forces that can be either enhanced or diminished compared to classical GR, depending on the specific parameter values \cite{isz24,isz25}. The trajectories naturally fall into well-established categories including bound orbits, plunging trajectories toward the BH, and scattering or escape paths, with the classification determined by the particle's energy and angular momentum. Remarkably, despite the complex underlying spacetime geometry, neutral particle motion maintains its regular and integrable character, contrasting sharply with the potentially chaotic dynamics exhibited by charged particles in electromagnetic fields \cite{isz26}.

We restrict our analysis to geodesic motion in the equatorial plane, defined by $\theta=\pi/2$ and $\dot{\theta}=0$. The Lagrangian density function $\mathcal{L}=\frac{1}{2}\,g_{\mu\nu}\,\dot{x}^{\mu}\,\dot{x}^{\nu}$ using the metric (\ref{aa1}) becomes:
\begin{equation}
    \mathcal{L}=\frac{1}{2}\,\left[-F(r)\,\dot{t}^2+\frac{\dot{r}^2}{F(r)}+r^2\,\dot{\phi}^2\right].\label{bb1}
\end{equation}

The spherical symmetry and time independence of our spacetime yield two conserved quantities associated with the Killing vector fields $\xi_{(t)} \equiv \partial_t$ and $\xi_{(\phi)} \equiv \partial_{\phi}$:
\begin{equation}
    \mathrm{E}=F(r)\,\dot{t}\quad,\quad \mathrm{L}=r^2\,\dot{\phi}.\label{bb2}
\end{equation}

Substituting Eq. (\ref{bb2}) into Eq. (\ref{bb1}), we obtain the fundamental geodesic equation:
\begin{equation}
    \dot{r}^2+V_\text{eff}(r)=\mathrm{E}^2,\label{bb3}
\end{equation}
where the effective potential encapsulates all geometric effects:
\begin{equation}
    V_\text{eff}(r)=\left(1+\frac{\mathrm{L}^2}{r^2}\right)\,F(r)=\left(1+\frac{\mathrm{L}^2}{r^2}\right)\,\left[\frac{1}{1 - \ell} - \frac{2\,M}{r} +\zeta\,\frac{e^{-\gamma}\,Q^2}{(1 - \ell)^2 r^2}\right].\label{bb4}
\end{equation}

\begin{figure}[ht!]
    \centering
    \includegraphics[width=0.3\linewidth]{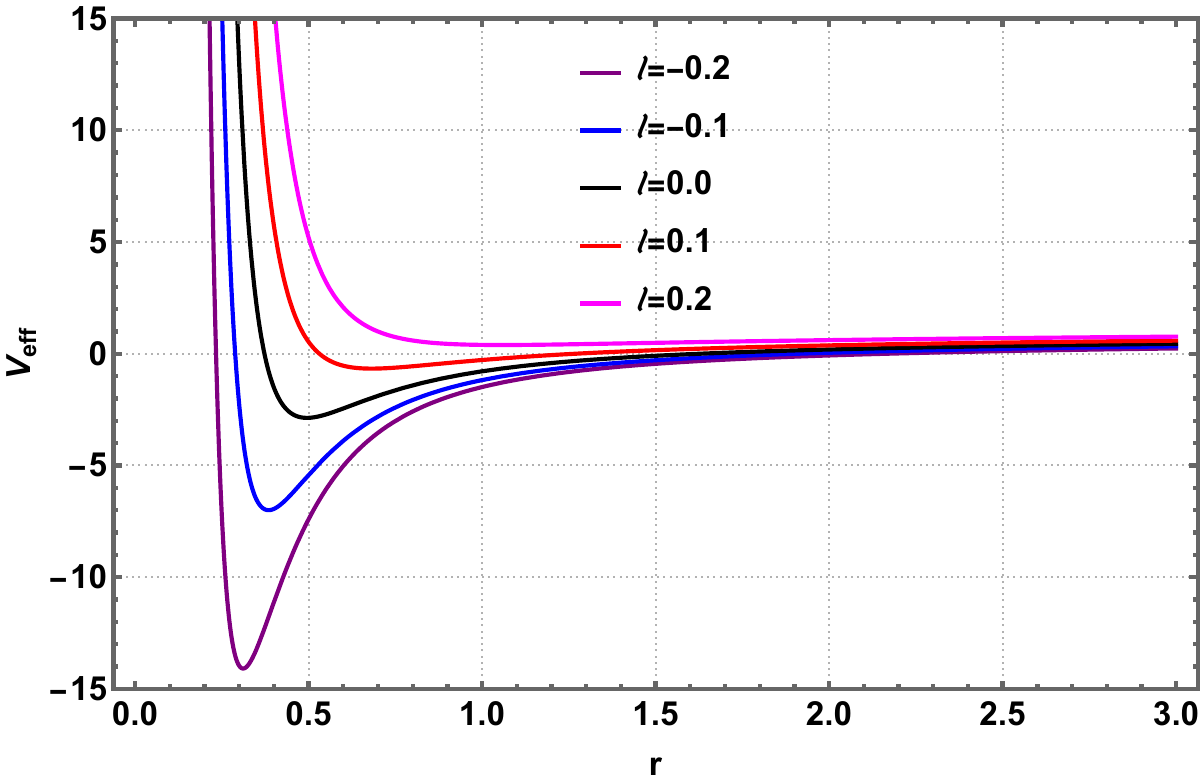}\quad\quad
    \includegraphics[width=0.3\linewidth]{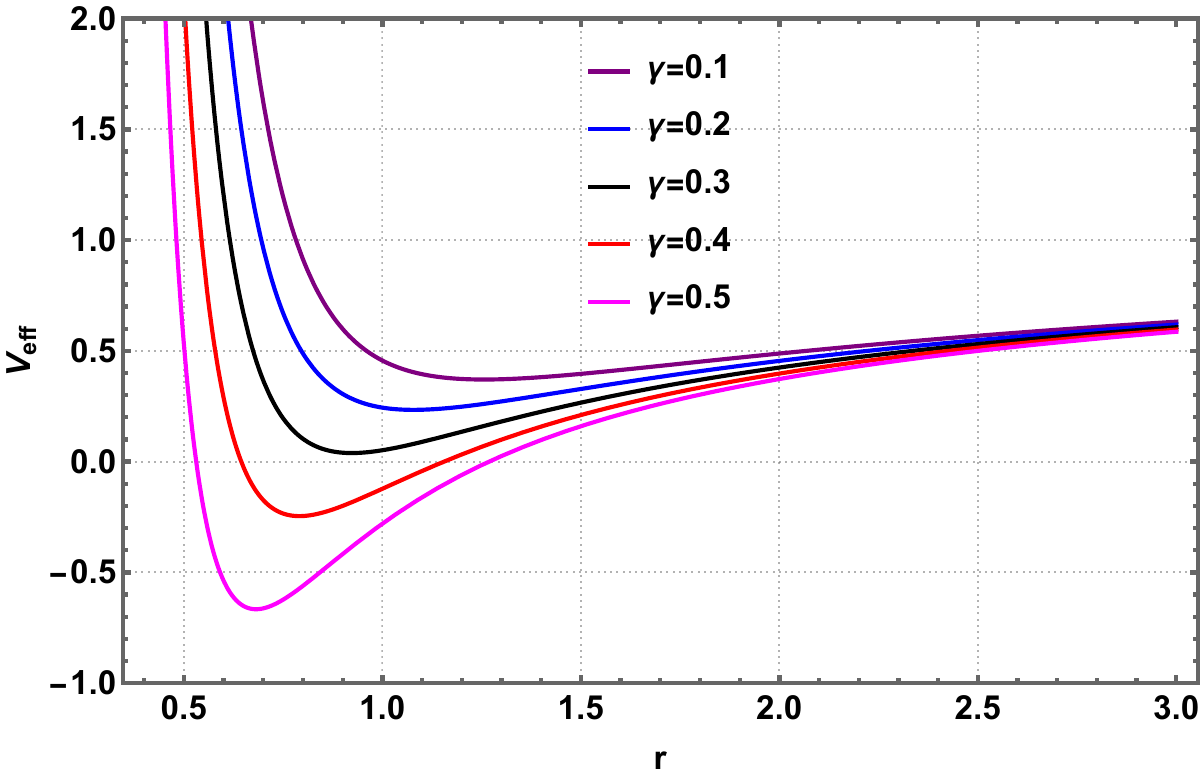}\quad\quad
    \includegraphics[width=0.3\linewidth]{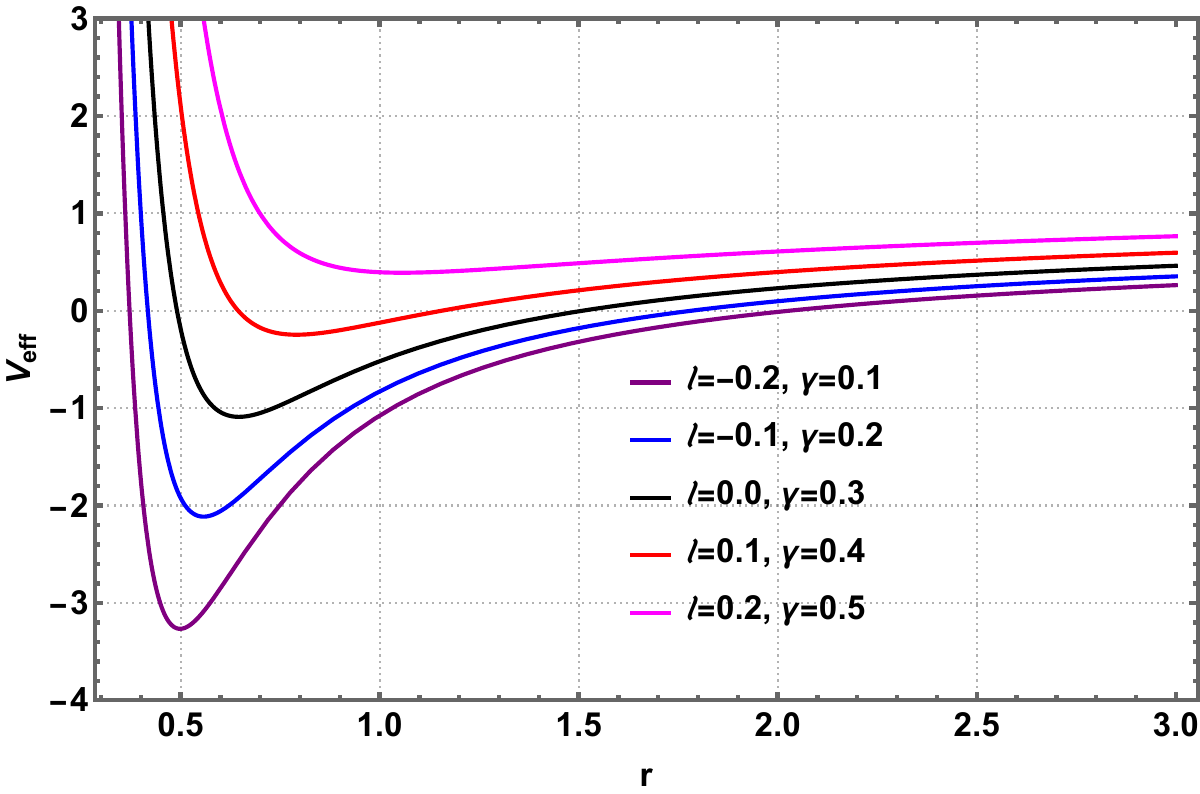}\\
    (a) $\gamma=0.5, Q=1$ \hspace{5cm} (b) $\ell=0.1 , Q=1$ \hspace{5cm} (c) $Q=1$\\
    \includegraphics[width=0.3\linewidth]{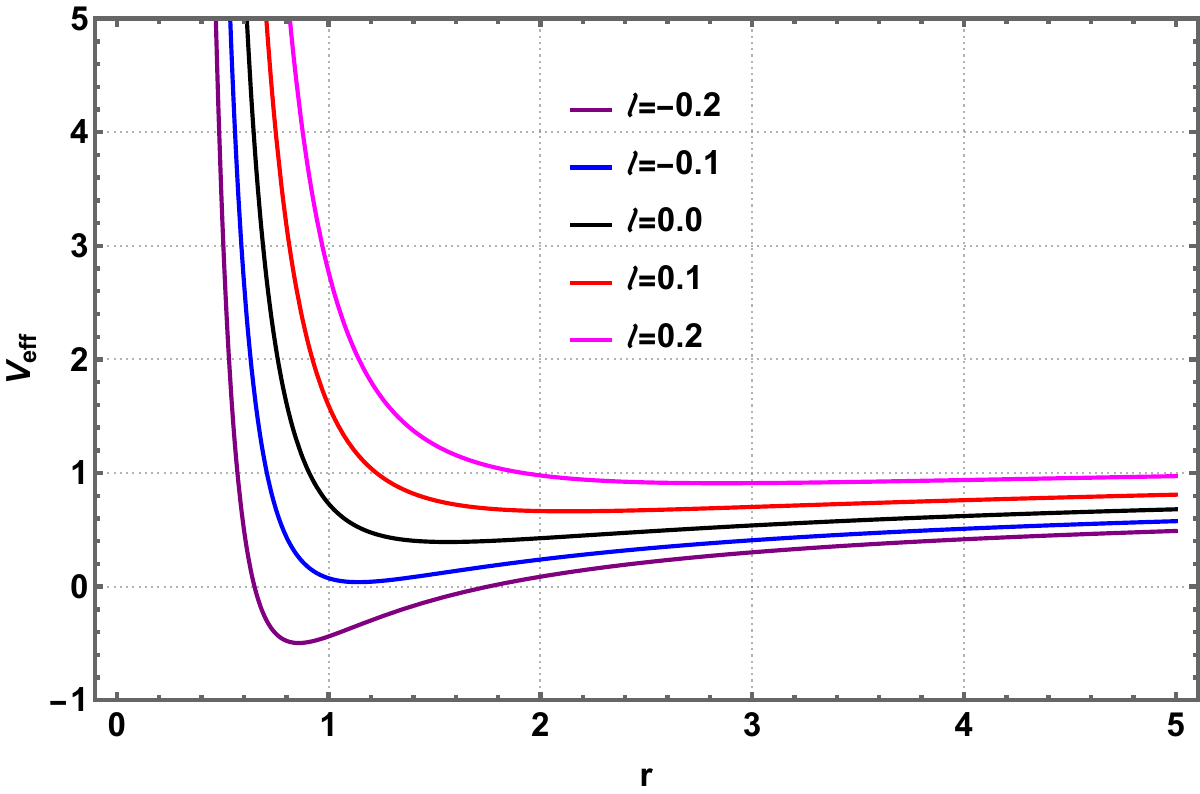}\quad\quad
    \includegraphics[width=0.3\linewidth]{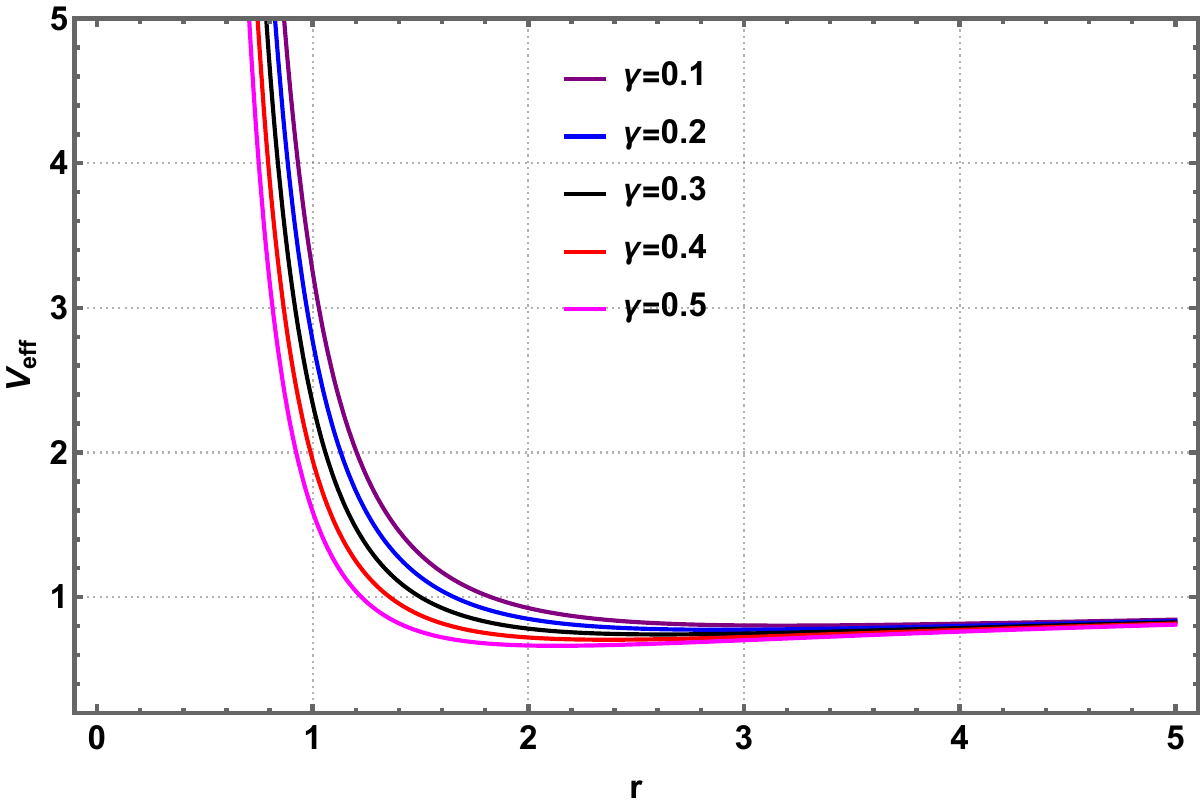}\quad\quad
    \includegraphics[width=0.3\linewidth]{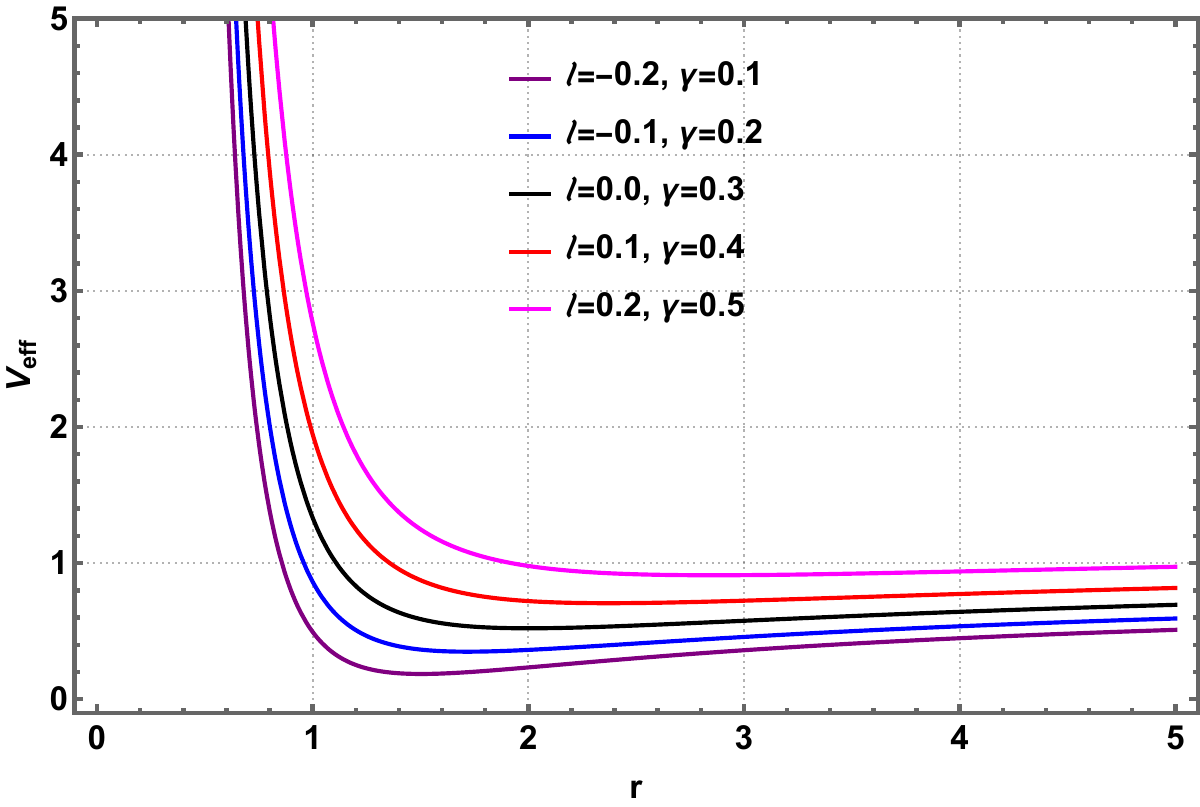}\\
    (d) $\gamma=0.5, Q=1.5$ \hspace{5cm} (e) $\ell=0.1 , Q=1.5$ \hspace{5cm} (f) $Q=1.5$\\
    \includegraphics[width=0.3\linewidth]{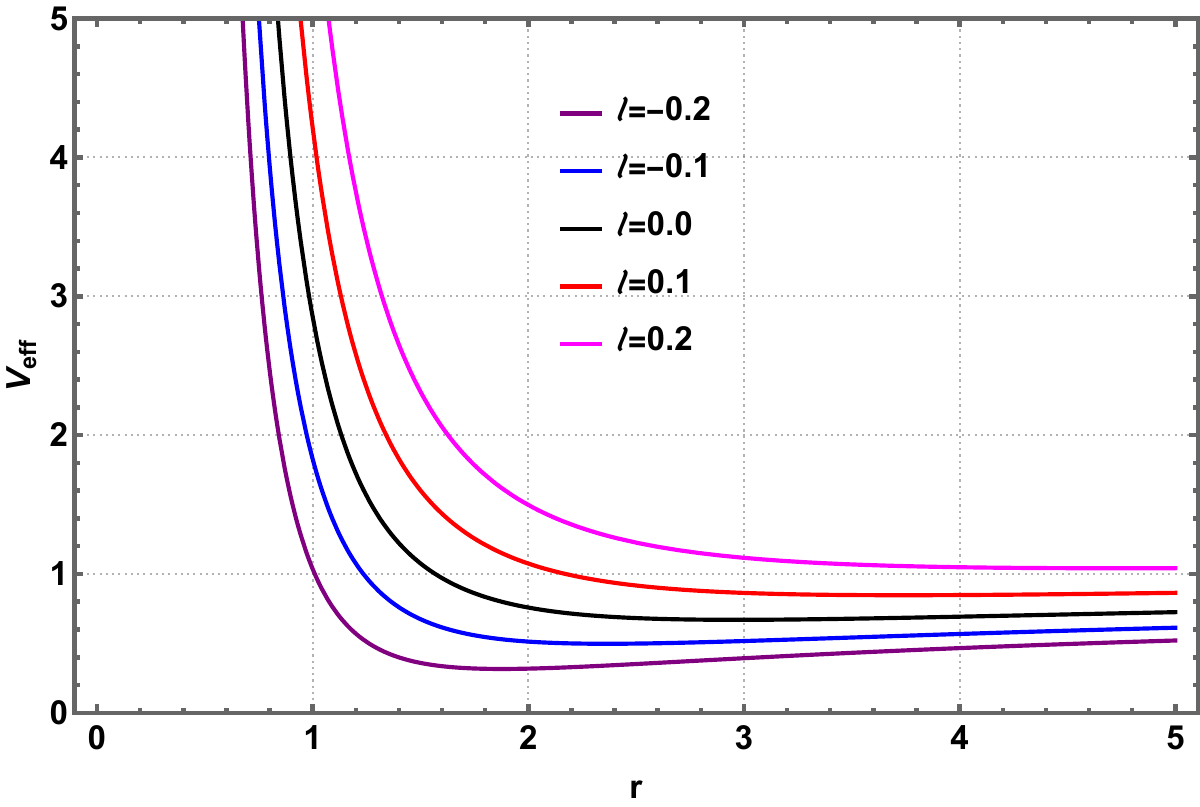}\quad\quad
    \includegraphics[width=0.3\linewidth]{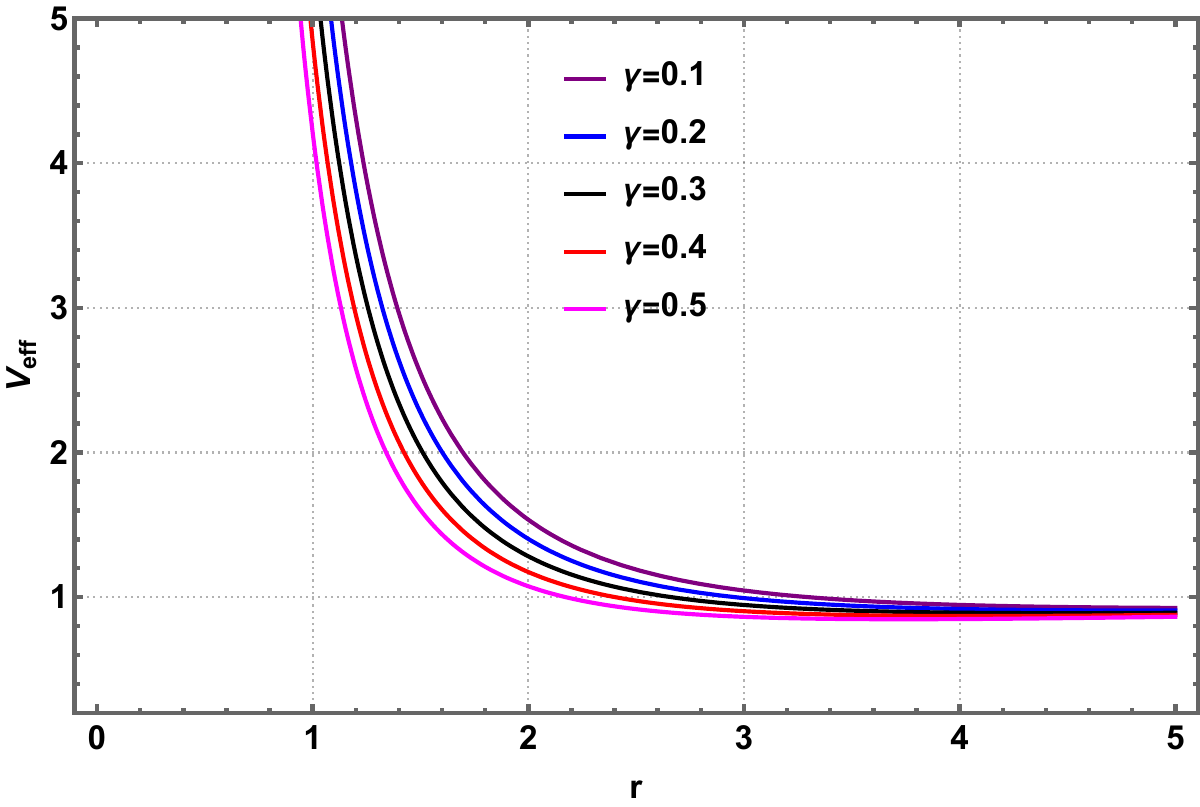}\quad\quad
    \includegraphics[width=0.3\linewidth]{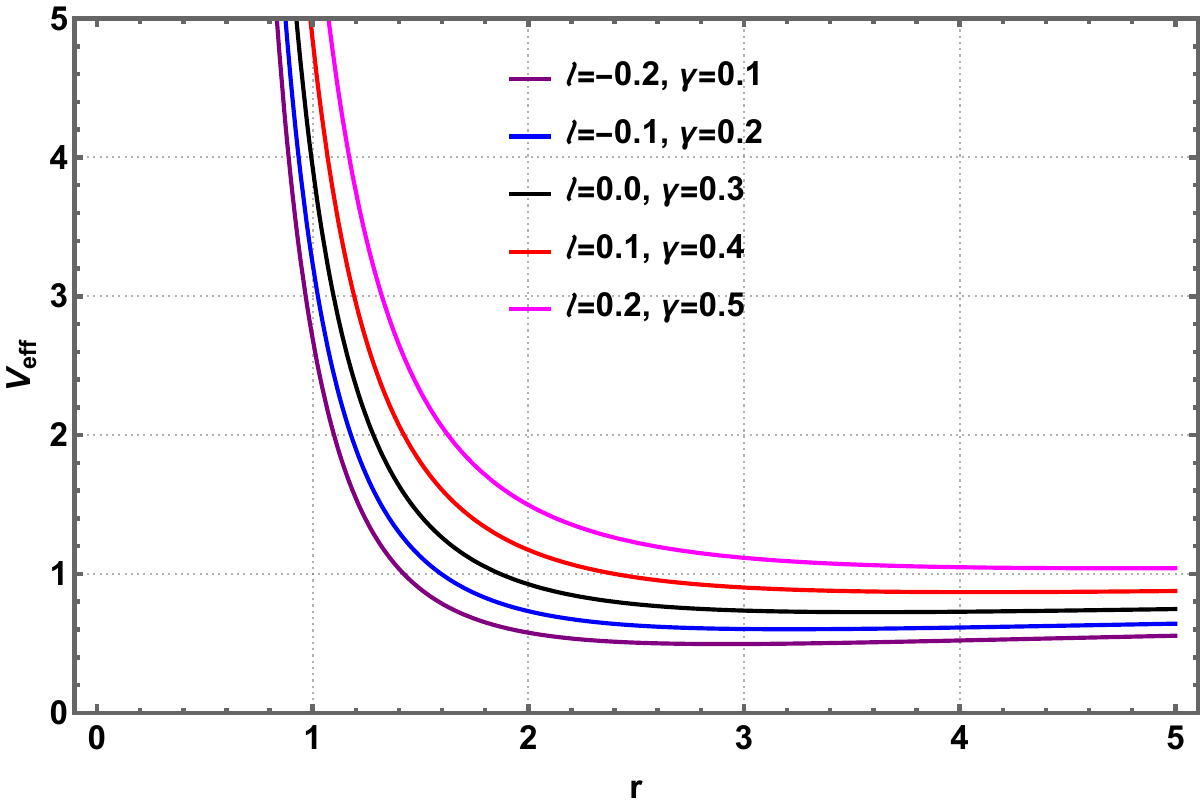}\\
    (g) $\gamma=0.5, Q=2$ \hspace{5cm} (h) $\ell=0.1 , Q=2$ \hspace{5cm} (i) $Q=2$\\
    \caption{\footnotesize Behavior of the effective potential $V_\text{eff}$ given in Eq. (\ref{bb4}) for different values of LSB parameter $\ell$ and ModMax parameter $\gamma$. Here, we set $M=1=\mathrm{L}$.}
    \label{fig:1}
\end{figure}

\begin{figure}[ht!]
    \centering
    \includegraphics[width=0.3\linewidth]{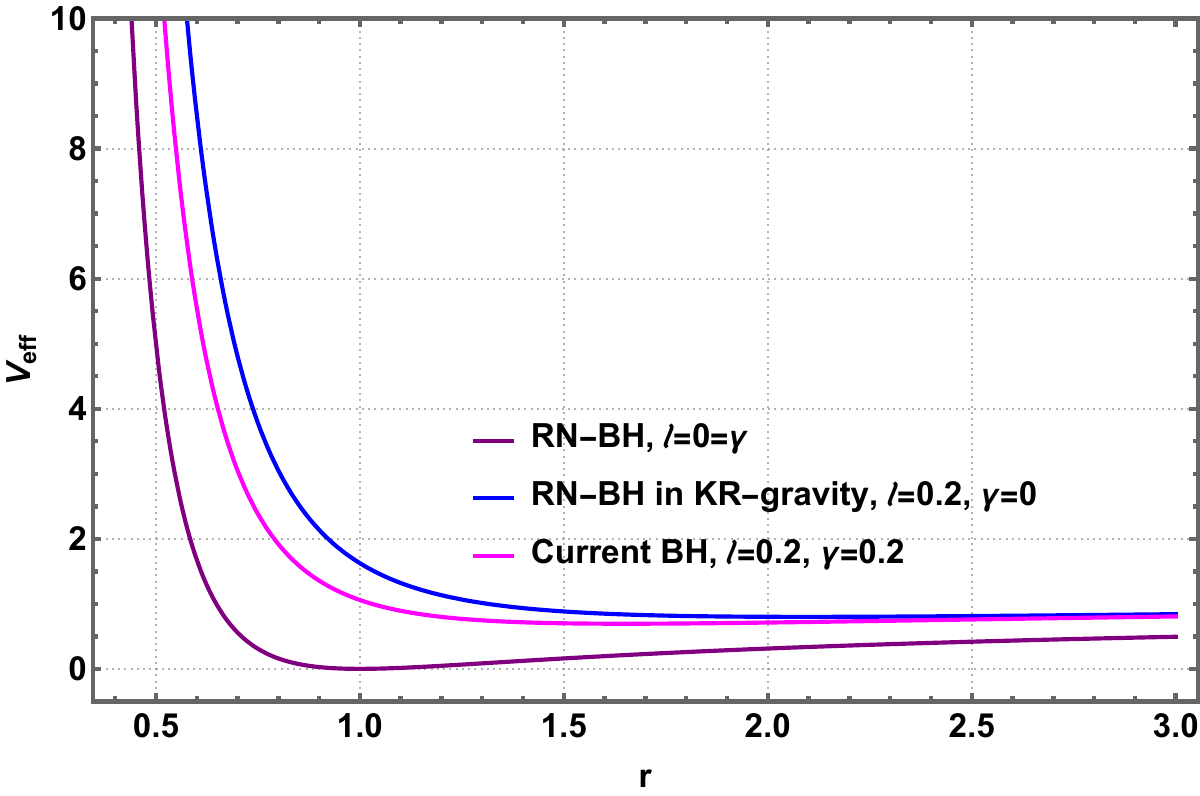}\quad\quad
    \includegraphics[width=0.3\linewidth]{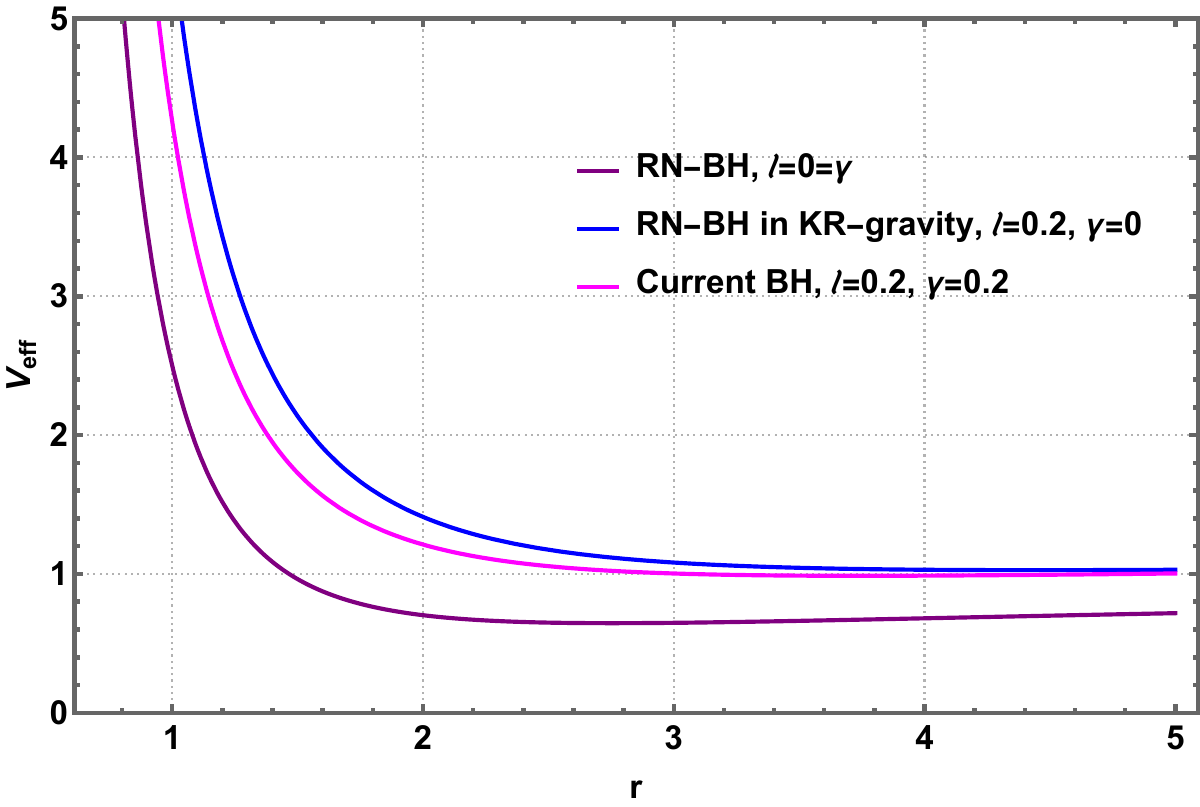}\quad\quad
    \includegraphics[width=0.3\linewidth]{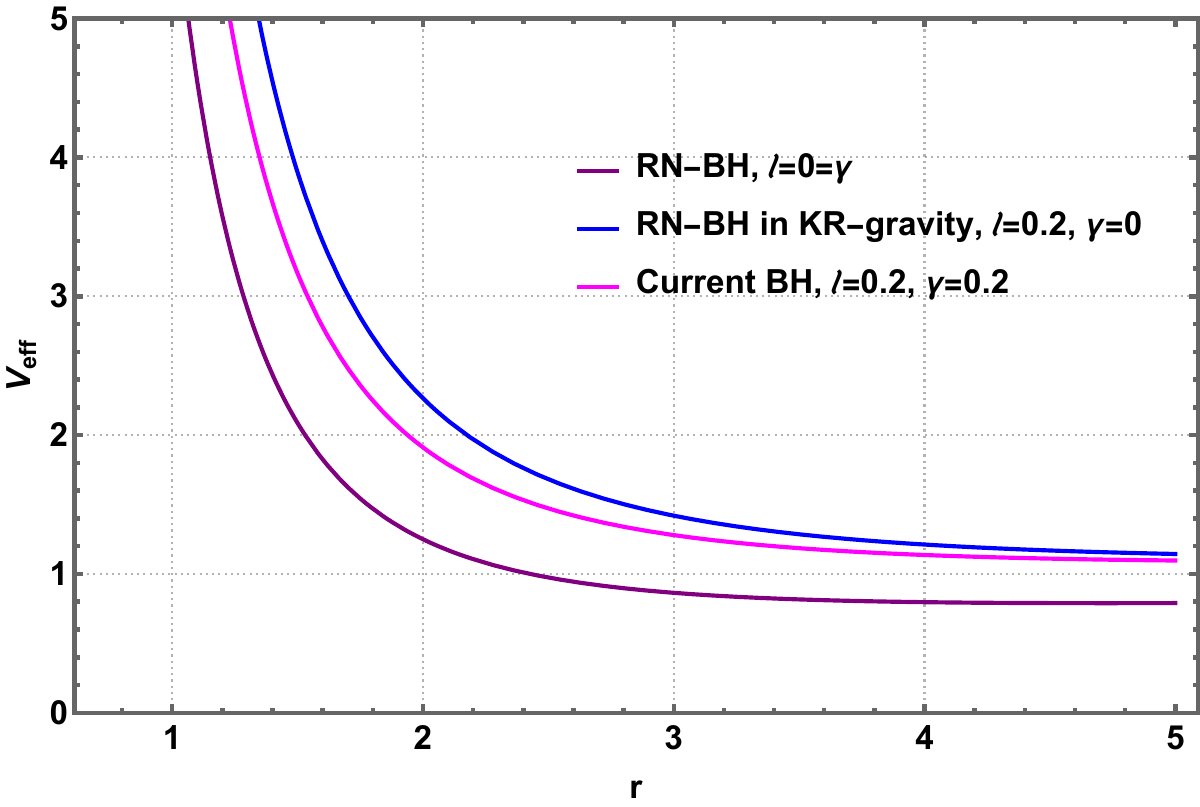}\\
    (a) $Q=1$ \hspace{5cm} (b) $Q=1.5$ \hspace{5cm} (c) $Q=2$\\
    \caption{\footnotesize Comparison of the effective potential $V_\text{eff}$ given in Eq. (\ref{bb4}) for different BH scenarios. Here, we set $M=1=\mathrm{L}$.}
    \label{fig:2}
\end{figure}

The comprehensive analysis of the effective potential presented in Figure \ref{fig:1} reveals the profound influence of LSB and ModMax parameters on neutral particle dynamics across different charge configurations. The systematic variation of the LSB parameter $\ell$ demonstrates how violations of Lorentz symmetry fundamentally alter the gravitational attraction, with negative values of $\ell$ generally enhancing the potential wells and positive values reducing them. The ModMax parameter $\gamma$ introduces subtle but significant modifications to the potential structure, particularly evident in the intermediate radial regions where electromagnetic effects become prominent. As the BH charge $Q$ increases from unity to two, the potential landscapes exhibit increasingly complex behavior, with the emergence of additional local extrema and modified barrier heights that directly impact orbital stability. The interplay between these parameters creates a rich phenomenology where traditional orbital classifications may break down, necessitating careful numerical analysis to determine particle trajectories. Figure \ref{fig:2} provides crucial comparative insights by contrasting the KR ModMax BH potential with classical RN and Schwarzschild cases, clearly demonstrating the substantial deviations introduced by the modified gravitational framework. The enhanced potential barriers and shifted turning points visible in these comparisons highlight the observational signatures that could distinguish this exotic BH from conventional solutions.

For circular orbital motion, the conditions $\dot{r}=0$ and $\ddot{r}=0$ must be simultaneously satisfied. The first condition yields:
\begin{equation}
    \mathrm{E}^2=\left(1+\frac{\mathrm{L}^2}{r^2}\right)\,F(r),\label{bb5}
\end{equation}

while the second condition determines the specific angular momentum:
\begin{equation}
    \mathrm{L}_\text{spc.}=r\,\sqrt{\frac{\frac{M}{r}-\zeta\,\frac{e^{-\gamma}\,Q^2}{(1-\ell)\,r^2}}{\frac{1}{1 - \ell} - \frac{3\,M}{r} + 2\,\zeta\, \frac{e^{-\gamma}\, Q^2}{(1 - \ell)^2\, r^2}}}.\label{bb6}
\end{equation}
Combining these results yields the specific energy:
\begin{equation}
    \mathrm{E}_\text{spc.}=\frac{\frac{1}{1 - \ell} - \frac{2\,M}{r} +\zeta\,\frac{e^{-\gamma}\,Q^2}{(1 - \ell)^2 r^2}}{\sqrt{\frac{1}{1 - \ell} - \frac{3\,M}{r} + 2\,\zeta\, \frac{e^{-\gamma}\, Q^2}{(1 - \ell)^2\, r^2}}}.\label{bb7}
\end{equation}
From the above expressions (\ref{bb6}) and (\ref{bb7}), it becomes clear that the specific angular momentum and specific energy of neutral test particles are influenced by geometrical and physical parameters. These include the Mod Max parameter $\gamma$, the electric charge $Q$, Lorentz-violating parameter $\ell$, and the BH mass $M$. Thus the results get modification in comparison to the standard BH solution.

The ISCO analysis requires satisfaction of three critical conditions:
\begin{align}
    \mathrm{E}^2 =V_{\text{eff}},\quad\quad V'_{\text{eff}}= 0,\quad\quad V''_{\text{eff}} \geq 0.\label{cond}
\end{align}

These conditions translate into a constraint on the metric function:
\begin{equation}
    F\,F'' + \frac{3\,F\,F'}{r} - 2\,{F'}^2=0,\label{cond2}
\end{equation}

leading to the polynomial equation for the ISCO radius:
\begin{equation}
    \frac{M}{(1 - \ell)}\,r^3 
- 6\,M^2\,r^2 
+ \frac{9\,M\, \zeta\, e^{-\gamma}\, Q^2}{(1 - \ell)^2}\,r 
- \frac{4\, \zeta^2\, e^{-2\gamma}\, Q^4}{(1 - \ell)^4}=0.\label{cond3}
\end{equation}

In the limiting case $Q=0$, this reduces to the Schwarzschild-KR result:
\begin{equation}
    r_\text{ISCO}=6\,M\,(1 - \ell)\label{cond4}
\end{equation}

\begin{table*}[ht!]
\centering
\scriptsize
\setlength{\tabcolsep}{4pt}
\renewcommand{\arraystretch}{2}
\begin{adjustbox}{max width=\textwidth}
\begin{tabular}{|c|ccc|ccc|ccc|ccc|}
\hline
 & \multicolumn{3}{c|}{$\ell=0.2$} & \multicolumn{3}{c|}{$\ell=0.3$} & \multicolumn{3}{c|}{$\ell=0.4$} & \multicolumn{3}{c|}{$\ell=0.5$} \\
\hline
$\gamma$ & 0.1 & 0.2 & 0.3 & 0.1 & 0.2 & 0.3 & 0.1 & 0.2 & 0.3 & 0.1 & 0.2 & 0.3 \\
\hline
$Q = 1$   & 1.0047 & 0.9363 & 0.8751 & 1.1843 & 1.0978 & 1.0184 & 1.4502 & 1.3419 & 1.2419 & 1.8503 & 1.7116 & 1.5833 \\
$Q = 1.5$ & 1.8569 & 1.7183 & 1.5905 & 2.2193 & 2.0529 & 1.8991 & 2.7261 & 2.5224 & 2.3337 & 3.4664 & 3.2098 & 2.9716 \\
$Q = 2$   & 2.9063 & 2.6885 & 2.4870 & 3.4715 & 3.2125 & 2.9725 & 4.2514 & 3.9368 & 3.6449 & 5.3784 & 4.9852 & 4.6199 \\
\hline
\end{tabular}
\end{adjustbox}
\caption{\footnotesize Numerical values of the ISCO radius $r=r_\text{ISCO}$ using Eq.~(\ref{cond3}) for different values of KR-field parameter $\ell$, ModMax parameter $\gamma$ and the electric charge $Q$. Here, we set $M=1$, $\zeta=1$.}
\label{tab:1}
\end{table*}

\begin{table*}[ht!]
\centering
\scriptsize
\setlength{\tabcolsep}{4pt}
\renewcommand{\arraystretch}{2}

\begin{adjustbox}{max width=\textwidth}
\begin{tabular}{|c|ccc|ccc|ccc|ccc|}
\hline
 & \multicolumn{3}{c|}{$\ell=0.2$} & \multicolumn{3}{c|}{$\ell=0.3$} & \multicolumn{3}{c|}{$\ell=0.4$} & \multicolumn{3}{c|}{$\ell=0.5$} \\
\hline
$\gamma$ & 0.1 & 0.2 & 0.3 & 0.1 & 0.2 & 0.3 & 0.1 & 0.2 & 0.3 & 0.1 & 0.2 & 0.3 \\
\hline
$Q = 1$   & 6.51357 & 6.37396 & 6.24458 & 6.29063 & 6.12647 & 5.97378 & 6.18889 & 5.99426 & 5.81257 & 6.26634 & 6.03244 & 5.81343 \\
$Q = 1.5$ & 8.11440 & 7.86401 & 7.63036 & 8.14329 & 7.85605 & 7.58736 & 8.35479 & 8.02145 & 7.70904 & 8.84148 & 8.44722 & 8.07725 \\
$Q = 2$   & 9.93516 & 9.56669 & 9.22173 & 10.2176 & 9.79933 & 9.40717 & 10.7497 & 10.2680 & 9.81593 & 11.6653 & 11.0981 & 10.5655 \\
\hline
\end{tabular}
\end{adjustbox}

\caption{\footnotesize Numerical values of the ISCO radius $r=r_\text{ISCO}$ using Eq.~(\ref{cond3}) for different values of KR-field parameter $\ell$, ModMax parameter $\gamma$ and the electric charge $Q$. Here, we set $M=1$, $\zeta=-1$.}
\label{tab:2}
\end{table*}

Tables \ref{tab:1} and \ref{tab:2} present a comprehensive numerical investigation of ISCO radii across the parameter space of the KR ModMax theory, revealing striking differences between ordinary and phantom branch configurations. The ordinary branch (Table \ref{tab:1}) exhibits ISCO radii that are systematically smaller than their phantom counterparts, indicating that stable circular orbits can exist much closer to the BH horizon in ordinary configurations. The consistent trend shows that increasing the ModMax parameter $\gamma$ leads to inward migration of the ISCO, while larger LSB parameters $\ell$ push the ISCO outward, demonstrating the competing effects of nonlinear electrodynamics and spacetime symmetry breaking. For the phantom branch (Table \ref{tab:2}), the ISCO radii are significantly larger, often by factors of 5-10, suggesting that these exotic BH configurations maintain stable particle orbits only at considerable distances from the central singularity. The charge dependence is particularly pronounced, with higher charges systematically increasing ISCO radii in both branches, reflecting the enhanced electromagnetic contributions to the effective gravitational potential. These results have profound implications for accretion disk physics and gravitational wave emission, as the location of the ISCO directly determines the efficiency of energy extraction and the characteristics of the transition from stable to plunging orbits \cite{sec3is06,sec3is07}.

\begin{figure}[ht!]  
\includegraphics[width=0.3\linewidth]{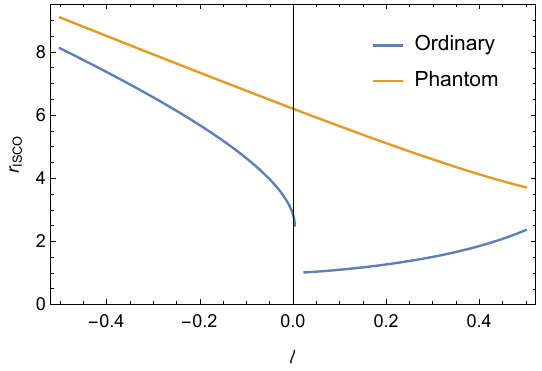}\quad\quad
\includegraphics[width=0.3\linewidth]{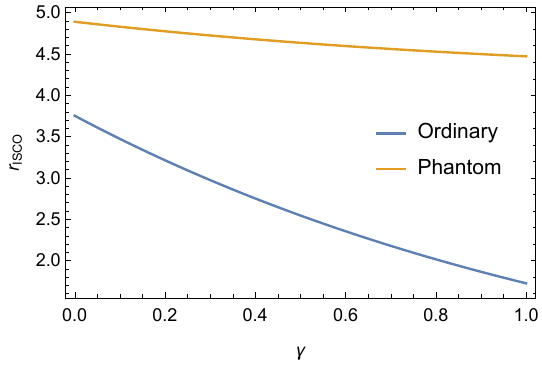}\quad\quad
\includegraphics[width=0.3\linewidth]{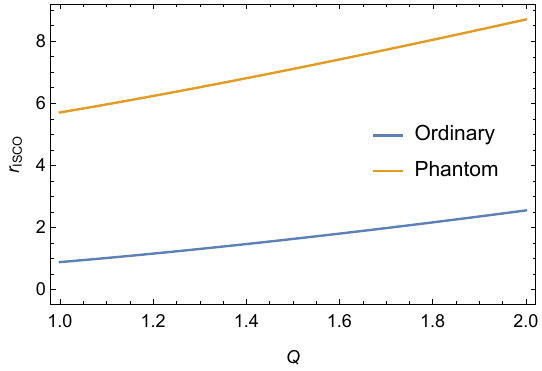}
\caption{\footnotesize Variation of the $r_{ISCO}$ for KR ModMax BH with various values of the parameters $\ell$ (left), $\gamma$ (middle) and $Q$ (right). Here, $M=1$.}
\label{figa19}
\end{figure}

Figure \ref{figa19} provides a comprehensive graphical analysis of ISCO behavior across the theory's parameter space, clearly illustrating the distinct characteristics of ordinary and phantom branches. The left panel demonstrates how the LSB parameter $\ell$ affects ISCO locations, with the phantom branch showing a dramatic increase in ISCO radius for positive $\ell$ values, while the ordinary branch exhibits more moderate variations. The middle panel reveals the systematic decrease in ISCO radius with increasing ModMax parameter $\gamma$ for both branches, though the effect is more pronounced in the ordinary configuration. The right panel shows the linear relationship between BH charge $Q$ and ISCO radius, with the phantom branch maintaining consistently larger values across all charge configurations. These trends collectively emphasize the critical role of LSB and ModMax parameters in determining orbital dynamics, with direct implications for observational signatures and tests of modified gravity theories.

The orbital dynamics are governed by the differential equation:
\begin{equation}
    \frac{d^2u}{d\phi^2}+\left(\frac{1}{1-\ell}+\zeta\,\frac{e^{-\gamma}\,Q^2}{(1 - \ell)^2\,\mathrm{L}^2}\right)\,u=\frac{M}{\mathrm{L}^2}+3\,M\,u^2-2\,\zeta\,\frac{e^{-\gamma}\,Q^2}{(1 - \ell)^2}\,u^3.\label{bb10}
\end{equation}

\begin{figure}[ht!]
    \centering
    \includegraphics[width=0.26\linewidth]{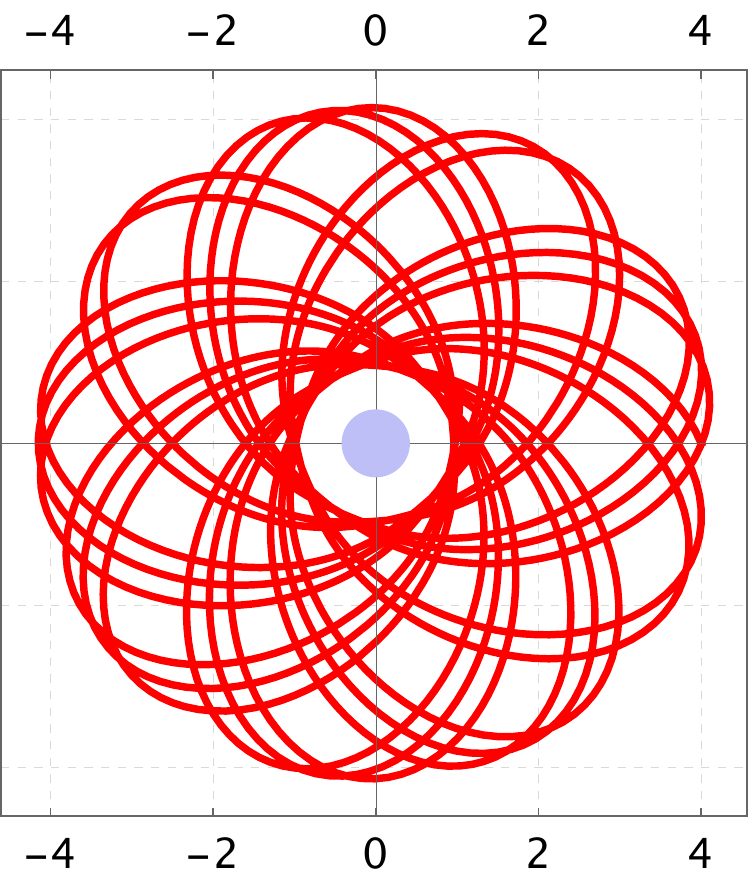}\quad\quad
    \includegraphics[width=0.26\linewidth]{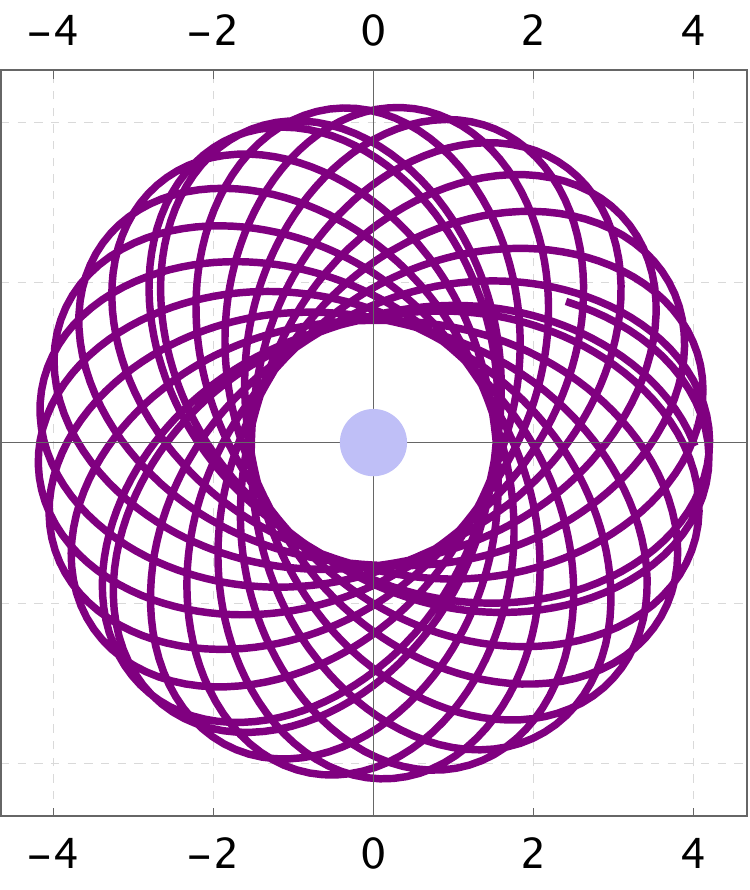}\quad\quad
    \includegraphics[width=0.26\linewidth]{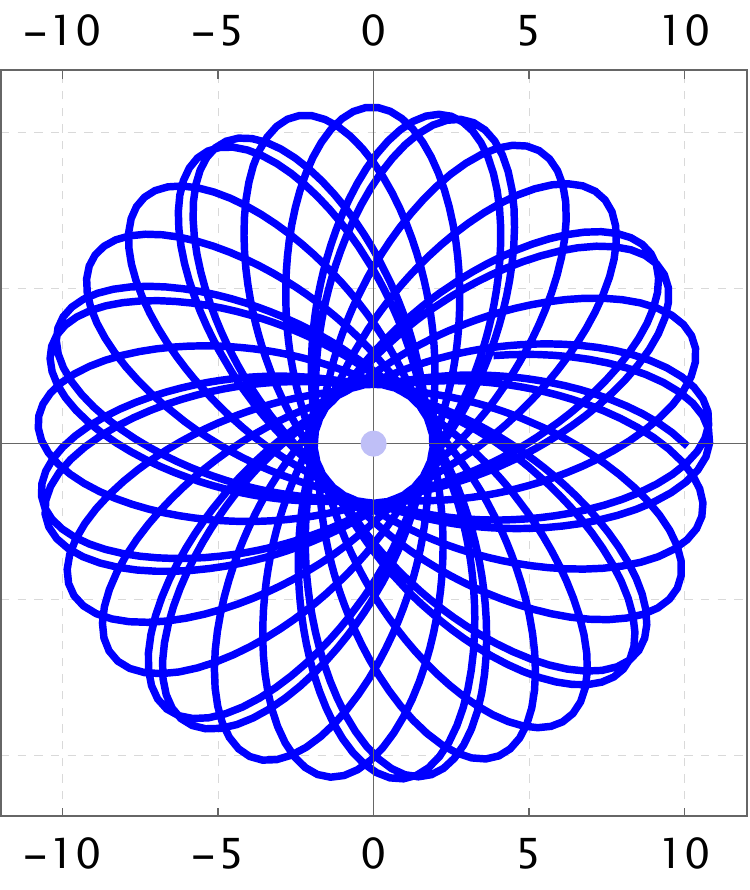}\\
    (a) $\ell=0.2$ \hspace{5cm} (b) $\ell=0.3$ \hspace{5cm} (c) $\ell=0.4$
    \caption{\footnotesize Illustration of neutral particle trajectories around an ordinary BH ($\zeta=1$) for different values of KR-field parameter $\ell$. Here, we set $M=1, \mathrm{L}=1, Q=1$, and $\gamma=0.3$.}
    \label{fig:trajectory-1}
    \includegraphics[width=0.26\linewidth]{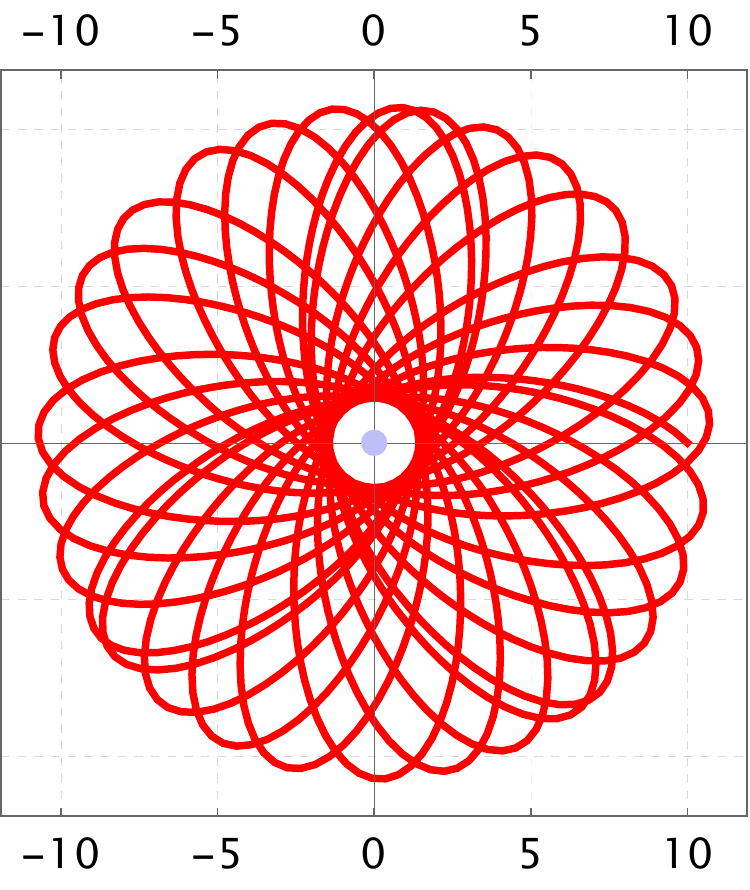}\quad\quad
    \includegraphics[width=0.26\linewidth]{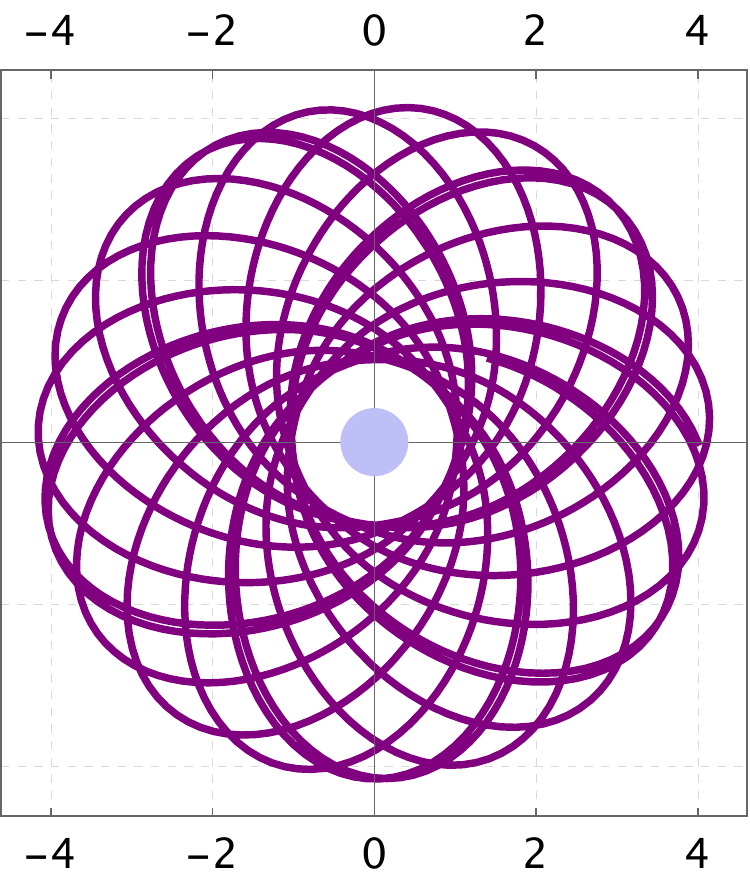}\quad\quad
    \includegraphics[width=0.26\linewidth]{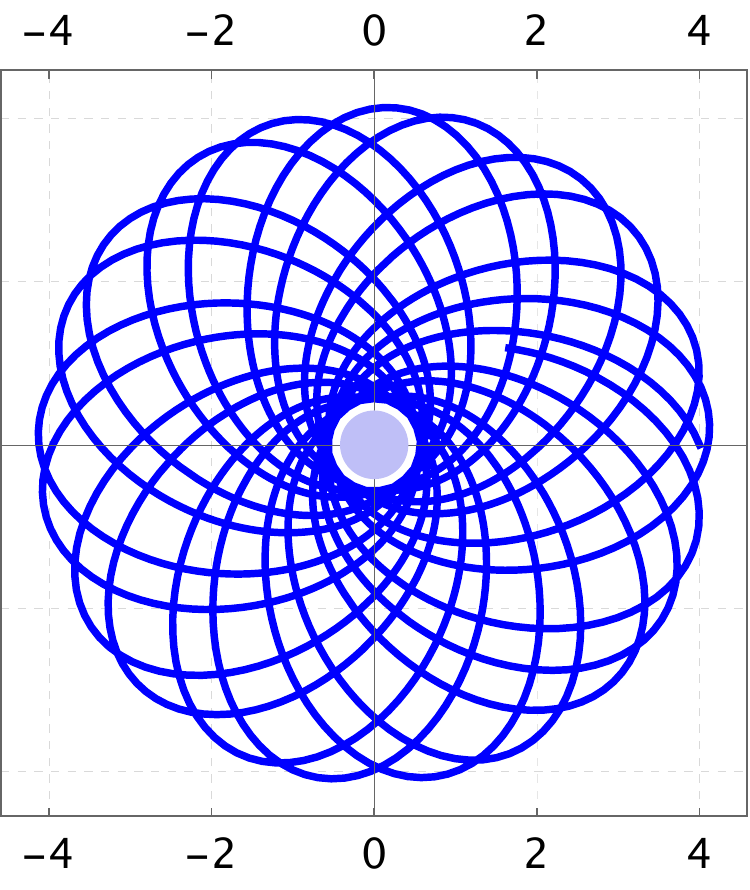}\\
    (d) $\gamma=0.2$ \hspace{5cm} (e) $\gamma=0.6$ \hspace{5cm} (f) $\gamma=1.0$\\
    \caption{\footnotesize Illustration of neutral particle trajectories around an ordinary BH ($\zeta=1$) for different values of ModMax parameter $\gamma$. Here, we set $M=1, \mathrm{L}=1, Q=1$ and $\ell=0.3$.}
    \label{fig:trajectory-2}
\end{figure}

\begin{figure}[ht!]
    \centering
    \includegraphics[width=0.26\linewidth]{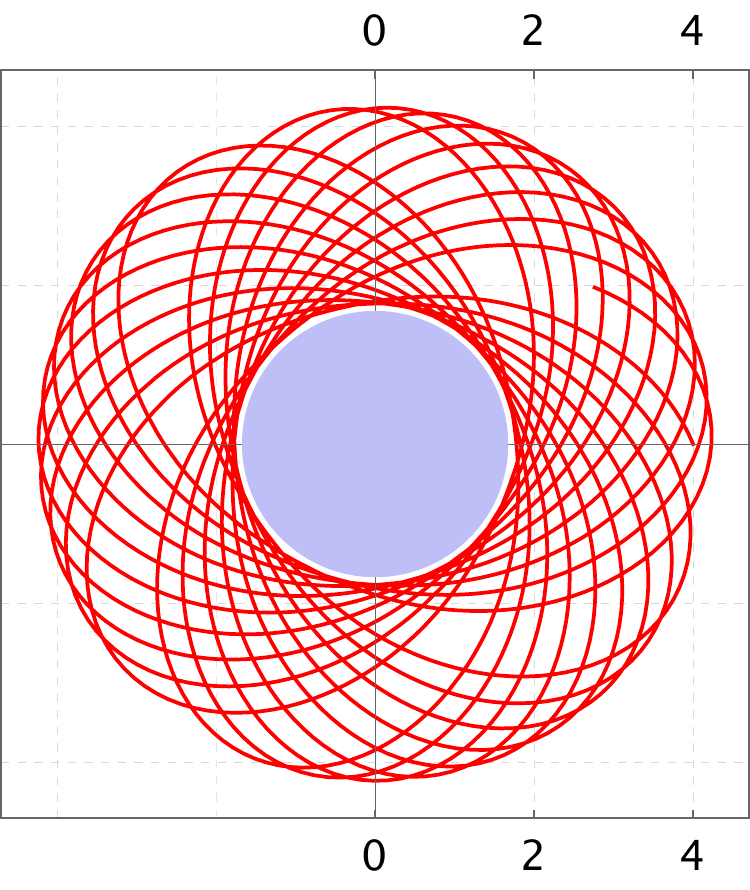}\quad\quad\quad\quad\quad
    \includegraphics[width=0.26\linewidth]{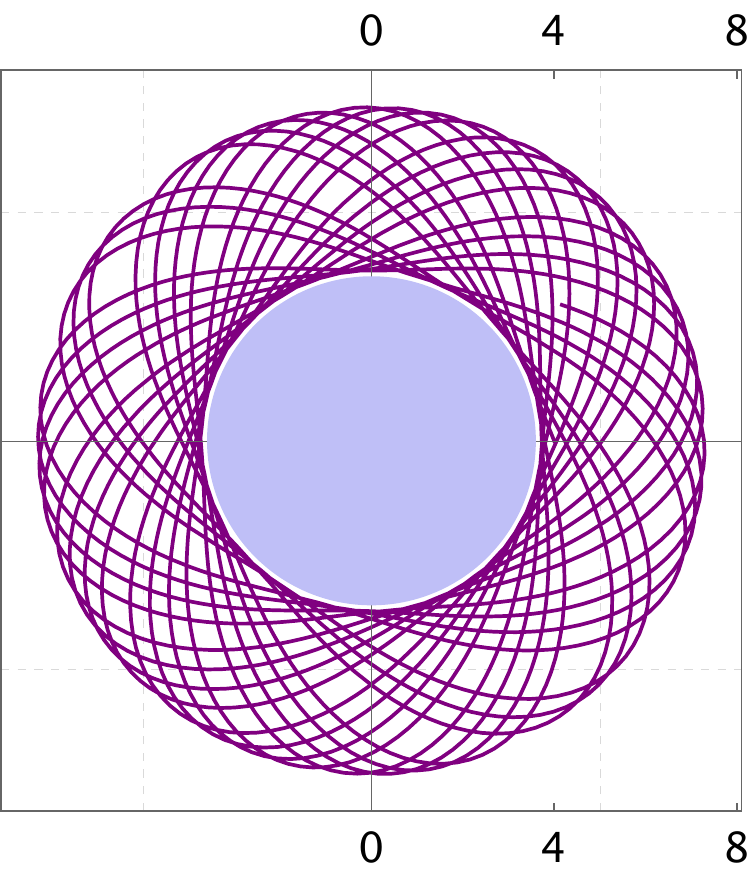}\\
    (a) $\ell=0.4,\,\gamma=0.6$ \hspace{6cm} (b) $\ell=0.6,\,\gamma=0.8$
    \caption{\footnotesize Illustration of neutral particle trajectories around an ordinary BH ($\zeta=1$) for combined values of ModMax parameter $\gamma$ and KR-field parameter $\ell$. Here, we set $M=1, \mathrm{L}=1, Q=1$.}
    \label{fig:trajectory-3}
\end{figure}

The trajectory visualizations in Figures \ref{fig:trajectory-1}, \ref{fig:trajectory-2}, and \ref{fig:trajectory-3} reveal the intricate orbital patterns that emerge from the interplay of LSB and ModMax effects in the KR ModMax BH spacetime. Figure \ref{fig:trajectory-1} demonstrates how varying the LSB parameter $\ell$ systematically alters the precession rates and overall trajectory shapes, with higher values leading to more tightly wound spirals and modified periapsis advance. The beautiful rosette patterns clearly show the departure from classical GR predictions, where such complex orbital structures would not be observed. Figure \ref{fig:trajectory-2} illustrates the influence of the ModMax parameter $\gamma$, showing how nonlinear electromagnetic effects contribute to orbital complexity through subtle modifications to the effective gravitational potential. The progression from simple elliptical-like orbits to more elaborate multi-looped structures demonstrates the rich phenomenology accessible within this theoretical framework. Figure \ref{fig:trajectory-3} presents the most sophisticated cases where both LSB and ModMax parameters are simultaneously active, resulting in highly complex orbital patterns that could serve as distinctive observational signatures for distinguishing these exotic BHs from their classical counterparts. These trajectory modifications would have direct observational consequences for stellar orbits around supermassive BHs and could potentially be detected through high-precision astrometric measurements.

Several important limiting cases illuminate the physical content of our general framework:

\begin{itemize}
    \item \textbf{Case 1: $Q = 0$} \\
    The uncharged limit reduces to Schwarzschild BH in KR gravity, with orbital dynamics governed by:
    \begin{equation}
    \frac{d^2u}{d\phi^2}+\frac{1}{1-\ell}\,u=\frac{M}{\mathrm{L}^2}+3\,M\,u^2.\label{bb11}
\end{equation}

    \item \textbf{Case 2: $\ell = 0$, $\zeta = 1$} \\
    This corresponds to pure ModMax electrodynamics without LSB:
    \begin{equation}
    \frac{d^2u}{d\phi^2}+\left(1+\frac{e^{-\gamma}\,Q^2}{\mathrm{L}^2}\right)\,u=\frac{M}{\mathrm{L}^2}+3\,M\,u^2-2\,e^{-\gamma}\,Q^2\,u^3.\label{bb12}
\end{equation}

    \item \textbf{Case 3: $\gamma = 0$, $\zeta = 1$} \\
    The standard RN BH in KR gravity scenario yields:
    \begin{equation}
    \frac{d^2u}{d\phi^2}+\left(\frac{1}{1-\ell}+\frac{Q^2}{(1 - \ell)^2\,\mathrm{L}^2}\right)\,u=\frac{M}{\mathrm{L}^2}+3\,M\,u^2-\frac{2\,Q^2}{(1 - \ell)^2}\,u^3.\label{bb13}
\end{equation}
\end{itemize}

\subsection{ Charged Particles} 

The dynamics of charged particles in the KR ModMax BH spacetime represent a fascinating and complex regime where gravitational, electromagnetic, LSB, and nonlinear electrodynamic effects converge to create rich phenomenological behavior. Unlike neutral particles that respond only to spacetime curvature, charged test particles experience the full interplay of these fundamental forces, leading to significantly more intricate trajectory patterns \cite{sec4is01,sec4is02}. The electromagnetic field generated by the BH is fundamentally modified by both LSB effects and ModMax nonlinearity, which not only alter the spacetime geometry but also deform the electromagnetic four-potential $A_\mu$, introducing nontrivial corrections that dramatically influence particle motion. These modifications create effective potentials that can exhibit multiple local extrema, leading to complex orbital structures that have no analogue in classical GR \cite{sec4is03,sec4is04}.

The sensitivity of charged particle trajectories to both the sign and magnitude of the particle charge creates a rich parameter space for exploration. Particles carrying the same electric charge as the BH experience electromagnetic repulsion that competes with gravitational attraction, typically pushing the ISCO outward and creating larger stable orbital radii \cite{sec4is05}. Conversely, oppositely charged particles are electromagnetically attracted toward the BH, allowing them to maintain stable orbits at smaller radii and potentially spiral closer to the event horizon before plunging. The presence of LSB and ModMax effects further modifies these electromagnetic interactions by altering both the field strength and spatial distribution, creating parameter-dependent shifts in orbital stability boundaries that could serve as observational signatures of exotic physics \cite{sec4is06,sec4is07}.

Perhaps most intriguingly, charged particle motion in this modified spacetime can transition between regular and chaotic dynamical regimes depending on the specific parameter values and initial conditions. While some regions of parameter space support stable, integrable motion reminiscent of classical systems, others exhibit sensitive dependence on initial conditions and complex, unpredictable trajectories. This chaotic behavior is particularly pronounced near critical regions such as the ISCO or photon sphere, where gravitational and electromagnetic forces achieve comparable magnitudes and their nonlinear interplay becomes dominant. The degree of chaos typically increases with the particle's charge-to-mass ratio, suggesting that highly charged particles would exhibit the most complex dynamics, with potential implications for high-energy astrophysical phenomena including particle acceleration mechanisms, jet formation processes, and electromagnetic emission from accretion flows \cite{sec4is08}.

The electromagnetic four-potential for the KR ModMax BH takes the covariant form:
\begin{equation}
A_\mu = \left(\mp\,\frac{e^{-\gamma/2}\, Q}{(1 - \ell)\, r},\, 0,\, 0,\, 0 \right),\label{cc1}
\end{equation}
where the sign depends on the branch parameter $\zeta=\pm 1$.

The equations of motion for charged test particles are derived from the generalized Lagrangian density function that incorporates electromagnetic coupling \cite{sec4is09}:
\begin{equation}
\mathcal{L} = \frac{1}{2} g_{\mu\nu} \dot{x}^\mu \dot{x}^\nu + q\, A_\mu\, \dot{x}^\mu,\label{cc3}
\end{equation}
where $q$ represents the charge-to-mass ratio of the test particle, $A_\mu$ is the electromagnetic potential, and $g_{\mu\nu}$ is the spacetime metric.

For our specific KR ModMax geometry, the Lagrangian reduces to:
\begin{equation}
    \mathcal{L}=\frac{1}{2}\,\left[-F(r)\,\dot{t}^2+\frac{\dot{r}^2}{F(r)}+r^2\,\dot{\phi}^2\right]+q\,A_t\,\dot{t}.\label{cc4}
\end{equation}

The system possesses two conserved quantities analogous to the neutral particle case, but now modified by electromagnetic interactions:
\begin{align}
    -\mathrm{E}&=g_{tt}\,\dot{t}-q\,A_t,\label{cc5}\\
    \mathrm{L}&=r^2\,\dot{\phi}.\label{cc6}
\end{align}

From the energy conservation equation (\ref{cc5}), we can express the time coordinate derivative as:
\begin{equation}
    \dot{t}=\frac{1}{F(r)}\,(\mathrm{E}-q\,A_t).\label{cc7}
\end{equation}

Substituting the conserved quantities into the Lagrangian yields the fundamental equation governing radial motion:
\begin{equation}
    \dot{r}^2=(\mathrm{E}-q\,A_t)^2-F(r)\,\left(1+\frac{\mathrm{L}^2}{r^2}\right).\label{cc8}
\end{equation}

For circular orbital motion, the conditions $\dot{r}=0$ and $\mathrm{E}=U(r)$ must be satisfied, where the effective potential for charged particles takes the form:
\begin{equation}
    U^{\pm}(r)=q\,A_t \pm\,\sqrt{F(r)\,\left(1+\frac{\mathrm{L}^2}{r^2}\right)}.\label{cc10}
\end{equation}

This effective potential exhibits a fundamental duality structure, consisting of electromagnetic (Coulomb-like) and gravitational contributions. The system possesses an inherent symmetry under the transformation $\frac{q\,e^{-\gamma}\,Q}{(1-\ell)} \to -\frac{q\,e^{-\gamma}\,Q}{(1-\ell)}$, which maps $U^{+} \to -U^{-}$ and $U^{-} \to -U^{+}$. For physical consistency, we focus on the $U^{+}$ branch, which corresponds to positive energy solutions and ensures proper particle dynamics in the asymptotic region \cite{sec4is01}.

The determination of ISCO locations for charged particles requires satisfaction of three critical stability conditions:
\begin{equation}
     \mathrm{E}=U(r),\quad \partial_r\,U(r)=0,\quad \partial_{rr} U(r) \geq 0.\label{cc11}
\end{equation}

The first derivative condition, after algebraic manipulation, yields:
\begin{equation}
    F'(r) \left( 1 + \frac{L^2}{r^2} \right) 
- \frac{2 F(r) L^2}{r^3} 
+ \frac{2 q e^{-\gamma/2} Q}{(1 - \ell) r^2} \cdot \sqrt{F(r)\left( 1 + \frac{L^2}{r^2} \right)} 
= 0.\label{cc12}
\end{equation}

This transcendental equation can be solved numerically to determine the angular momentum $\mathrm{L}$ as a function of the orbital radius and physical parameters.

The second derivative condition, ensuring orbital stability, leads to a complex expression involving the metric function and its derivatives:
\begin{align}
q\, A_t''(r) + \frac{ 
2 F(r) \left(1 + \frac{L^2}{r^2}\right)^2 F''(r) 
- F'(r)^2 \left(1 + \frac{L^2}{r^2}\right)^2 
- \frac{4 L^2 F(r) \left(1 + \frac{L^2}{r^2}\right) F'(r)}{r^3} 
+ \frac{12 L^2 F(r)^2 \left(1 + \frac{L^2}{r^2}\right)}{r^4} 
- \frac{4 L^4 F(r)^2}{r^6}
}{
4 \left[ F(r) \left(1 + \frac{L^2}{r^2} \right) \right]^{3/2}
}=0.\label{cc13}
\end{align}

The complexity of this expression, involving both the electromagnetic potential and metric function derivatives, makes analytical solution extremely challenging. However, numerical approaches can provide accurate ISCO determinations across the parameter space, revealing how LSB and ModMax effects modify charged particle orbital stability compared to classical RN BHs. The resulting ISCO locations depend sensitively on the relative signs and magnitudes of the particle and BH charges, with same-sign configurations typically exhibiting larger ISCO radii due to electromagnetic repulsion, while opposite-sign configurations allow closer stable orbits through electromagnetic attraction.

\section{Thermal properties of KR ModMax BH} \label{isec5}

The thermodynamic analysis of the KR ModMax BH provides crucial insights into how LSB effects and nonlinear electrodynamics fundamentally alter the thermal behavior of exotic spacetimes. This investigation represents a comprehensive exploration of BH thermodynamics in modified gravity theories, where the interplay between KR field effects and ModMax nonlinearity creates rich phenomenological behavior absent in classical GR \cite{isz28,isz29}. Our analysis focuses on the derivation and detailed examination of key thermodynamic quantities including Hawking temperature, entropy, specific heat capacity, and Helmholtz free energy, with particular emphasis on understanding the distinct thermal signatures of ordinary and phantom branch configurations \cite{isz30,isz44}. The systematic comparison between these branches reveals fundamental differences in thermal stability, phase transition behavior, and global thermodynamic properties that could serve as observational discriminators for exotic BH solutions. Furthermore, the modifications introduced by LSB and ModMax parameters create parameter-dependent thermal behaviors that provide new avenues for testing fundamental physics through precision measurements of BH thermal radiation \cite{sec5is05,sec5is06}.

We parameterize our thermodynamic analysis in terms of the horizon radius $r=r_+$, which serves as the natural thermodynamic coordinate for BH systems. The mass of the KR ModMax BH can be expressed by solving the horizon condition $F(r_+)=0$:
\begin{equation}
    M=\frac{(1-\ell)r^2_++\zeta\,Q^2\,e^{-\gamma}}{2(1-\ell)^2r_+}.\label{temp21}
\end{equation}

This general expression encompasses several important limiting cases that illuminate the physical content of our theory. In the limit $\gamma=0=Q$, corresponding to the complete absence of ModMax electrodynamics, the BH mass reduces to the pure KR gravity result:
\begin{equation}
   M=\frac{r_+}{2(\ell-1)}.\label{Mass2}
\end{equation}

Conversely, when $\ell=0$ (no LSB effects), we recover the ModMax electrodynamics result without gravitational modifications \cite{sec5is07}:
\begin{equation}
   M=\frac{r^2_++\zeta\,Q^2\,e^{-\gamma}}{2r_+}.\label{Mass3}
\end{equation}

The Hawking temperature, representing the thermal radiation emitted by the BH, follows from the standard relation $T_H=\frac{F'(r_{+})}{4\,\pi}$:
\begin{equation}
T_H=\frac{(1-\ell)r^2_++\zeta\,Q^2\,e^{-\gamma}}{4\,\pi(1-\ell)^2r^3_+}.  \label{Temperature}
\end{equation}

For phantom branch configurations ($\zeta=-1$), temperature positivity requires the constraint $(1-\ell)r^2_+>Q^2\,e^{-\gamma}$, while ordinary branch configurations ($\zeta=1$) automatically satisfy temperature positivity conditions. Notably, the phantom branch consistently exhibits higher temperatures than the ordinary branch, as quantified by:
\begin{equation}
T_H (\zeta=-1)-T_H(\zeta=1)=\frac{Q^2\,e^{-\gamma}}{2\,\pi(1-\ell)^2r^3_+}>0.\label{temperature4}    
\end{equation}

\begin{figure}[ht!]  
\includegraphics[width=0.4\linewidth]{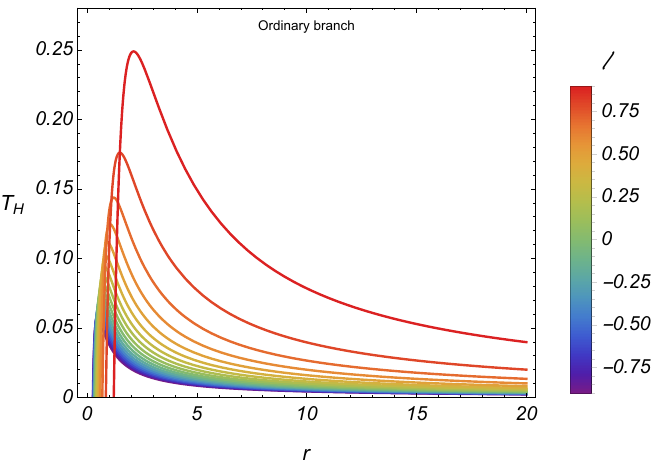}\quad\quad
\includegraphics[width=0.4\linewidth]{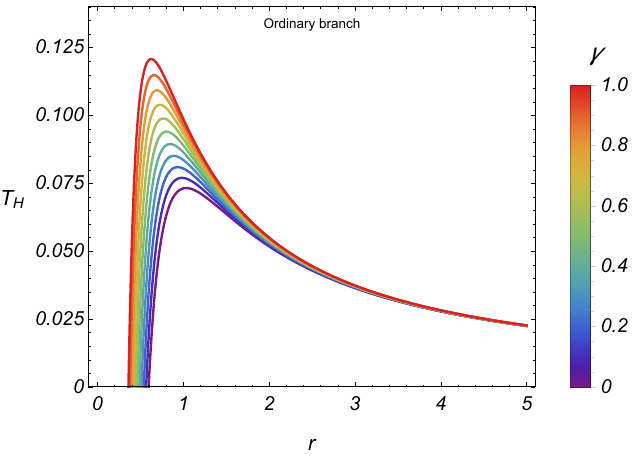}
\caption{\footnotesize The behavior of the Hawking temperature for KR ModMax BH with various values of the parameters $\ell$ (left) and $\gamma$ (right). Here, $M=1$ and $Q=0.5$.}
\label{figa3}
\end{figure}

\begin{figure}[ht!]    
\includegraphics[width=0.4\linewidth]{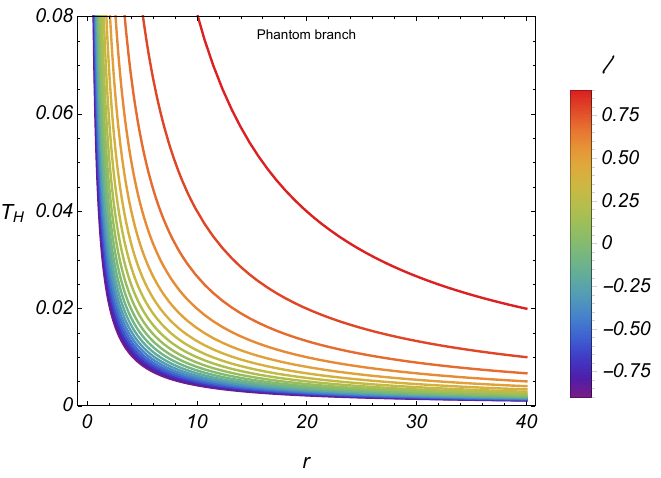}\quad\quad
\includegraphics[width=0.4\linewidth]{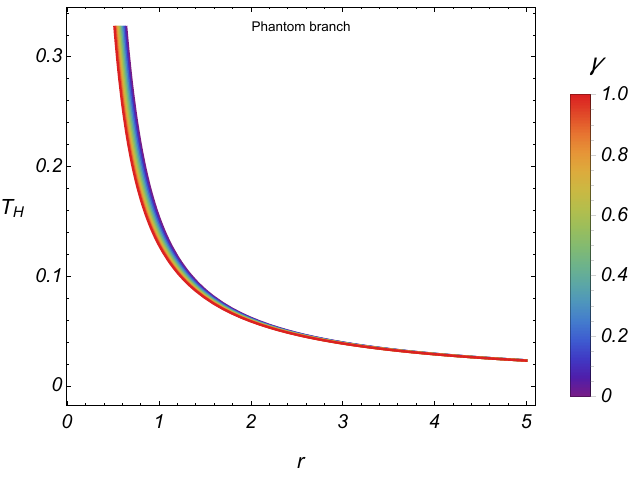}
\caption{\footnotesize The behavior of the Hawking temperature for KR ModMax BH with various values of the parameters $\ell$ (left) and $\gamma$ (right). Here, $M=1$ and $Q=0.5$.}
\label{figa4}
\end{figure}

Figures \ref{figa3} and \ref{figa4} provide comprehensive visualization of how the fundamental parameters $\ell$ and $\gamma$ influence Hawking temperature behavior across ordinary and phantom branch configurations. The ordinary branch (Figure \ref{figa3}) demonstrates that increasing the LSB parameter $\ell$ systematically raises the temperature across all horizon radii, reflecting how violations of Lorentz symmetry enhance the BH's thermal emission rate. The ModMax parameter $\gamma$ exhibits more subtle effects in the ordinary branch, with moderate increases generally leading to higher temperatures through modifications to the electromagnetic field structure. In stark contrast, the phantom branch (Figure \ref{figa4}) displays dramatically different temperature profiles, with the LSB parameter $\ell$ producing even more pronounced temperature enhancements that can exceed ordinary branch values by substantial factors. The ModMax parameter behavior in the phantom branch shows interesting non-monotonic patterns, where increasing $\gamma$ can initially suppress temperature before eventually enhancing it at larger horizon radii. These contrasting behaviors between ordinary and phantom branches provide clear observational signatures that could potentially distinguish between different theoretical scenarios through precision measurements of thermal radiation spectra. The temperature differences become particularly pronounced in the small horizon radius regime, where quantum effects are strongest and observational discrimination would be most feasible.

The entropy of the KR ModMax BH follows the standard area law, calculated using the fundamental thermodynamic relations:
\begin{equation}
S=\int \frac{1}{T } \frac{\partial {M}}{\partial r_{+}}dr_{+}=\pi\,r_+^2.
\label{entr2}
\end{equation} 

This result demonstrates that despite the complex modifications introduced by LSB and ModMax effects, the KR ModMax BH preserves the universal Bekenstein-Hawking entropy-area relationship, suggesting deep connections to the underlying statistical mechanical structure of quantum gravity \cite{sec5is08}.

To assess global thermodynamic stability, we analyze the Helmholtz free energy $G=M-T\,S$, which serves as a critical indicator of phase transition behavior:
\begin{equation}
    G=\frac{(1-\ell)r^2_++3\,\zeta\,Q^2\,e^{-\gamma}}{4(1-\ell)^2r_+}.\label{gibs2}
\end{equation}

\begin{figure}[ht!]  
\includegraphics[width=0.4\linewidth]{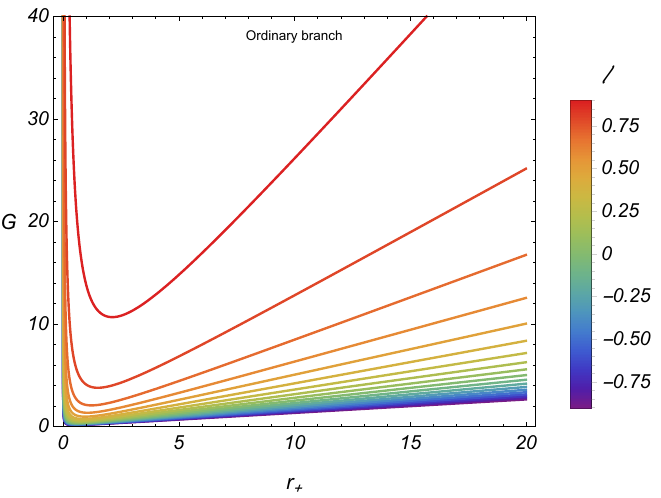}\quad\quad
\includegraphics[width=0.4\linewidth]{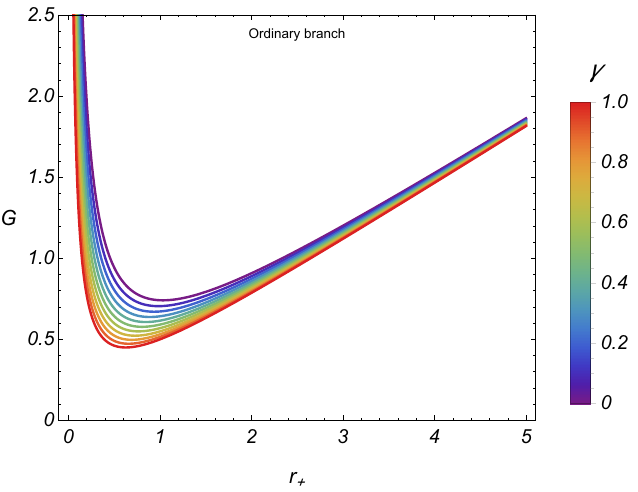}
\caption{\footnotesize The behavior of The Helmholtz energy of the KR ModMax BH with various values of the parameters $\ell$ (left) and $\gamma$ (right). Here, $M=1$ and $Q=0.5$.}
\label{figa5}
\end{figure}

\begin{figure}[ht!]    
\includegraphics[width=0.4\linewidth]{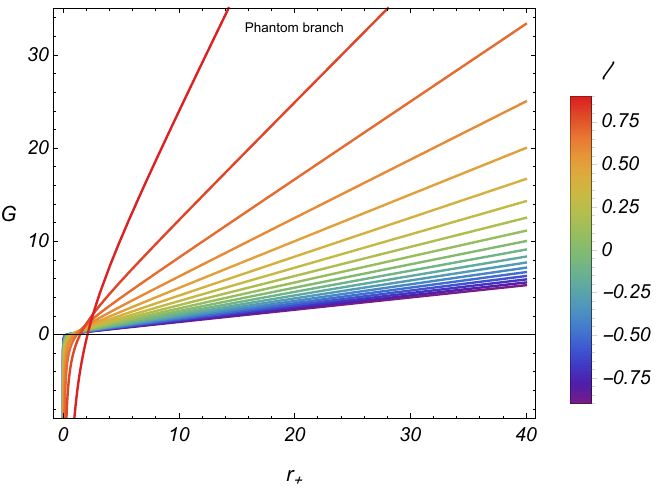}\quad\quad
\includegraphics[width=0.4\linewidth]{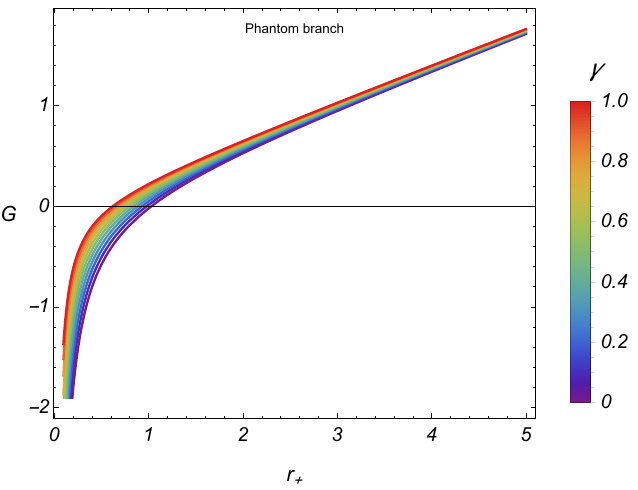}
\caption{\footnotesize The behavior of the Helmholtz energy of the KR ModMax BH with various values of the parameters $\ell$ (left) and $\gamma$ (right). Here, $M=1$ and $Q=0.5$.}
\label{figa6}
\end{figure} 

The Helmholtz free energy analysis presented in Figures \ref{figa5} and \ref{figa6} reveals fundamental differences in global thermodynamic stability between ordinary and phantom branch configurations. The ordinary branch (Figure \ref{figa5}) maintains consistently positive free energy values across the entire parameter range, indicating global thermodynamic instability without any Hawking-Page phase transitions. This behavior suggests that ordinary branch KR ModMax BHs would be thermodynamically disfavored compared to thermal radiation at the same temperature, making them inherently unstable systems. In contrast, the phantom branch (Figure \ref{figa6}) exhibits rich phase structure with clear transitions from negative to positive free energy regions, indicating the presence of Hawking-Page phase transitions that could stabilize these exotic BH configurations. The critical points where the free energy vanishes mark phase boundaries between stable and unstable thermodynamic regimes, providing potential observational targets for detecting phantom branch physics. The parameter dependence shows that LSB effects generally increase the free energy in both branches, while ModMax parameters can either enhance or suppress the free energy depending on the specific branch and parameter regime. These thermodynamic phase structures have profound implications for BH formation and evolution scenarios, as phantom branch configurations could potentially achieve thermal equilibrium with surrounding radiation fields under specific astrophysical conditions.

The specific heat capacity $C_+=\frac{dM}{dT_H}$ provides crucial information about thermal stability and phase transition behavior:
\begin{equation}
   C_+=\frac{4\,\pi(1-\ell)^2r^4_+}{3\,\zeta\,Q^2\,e^{-\gamma}-(1-\ell)r_+^2}.\label{heat}
\end{equation}

\begin{figure}[ht!]  
\includegraphics[width=0.4\linewidth]{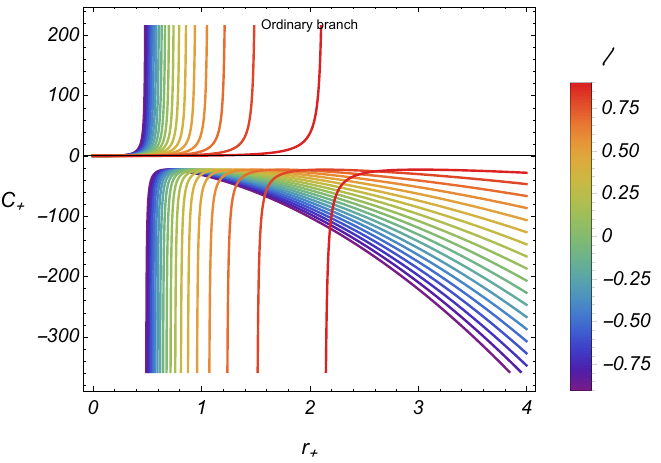}\quad\quad
\includegraphics[width=0.4\linewidth]{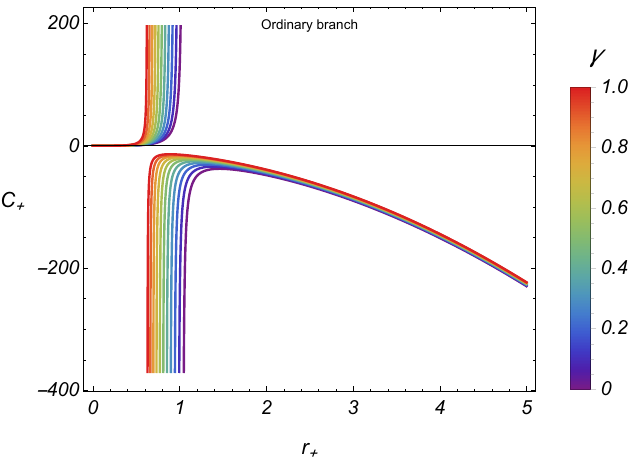}
\caption{\footnotesize Variation of the specific heat capacity $C_+$ as a function of horizon radius $r_h$ for KR ModMax BH with various values of the parameters $\ell$ (left) and $\gamma$ (right). Here, $M=1$ and $Q=0.5$.}
\label{figa7}
\end{figure}

\begin{figure}[ht!]    
\includegraphics[width=0.4\linewidth]{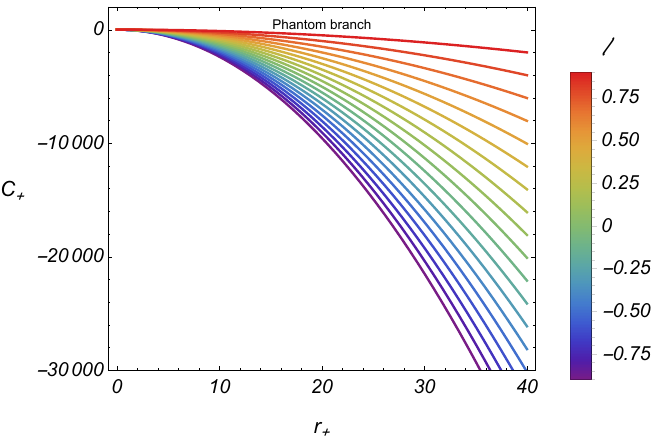}\quad\quad
\includegraphics[width=0.4\linewidth]{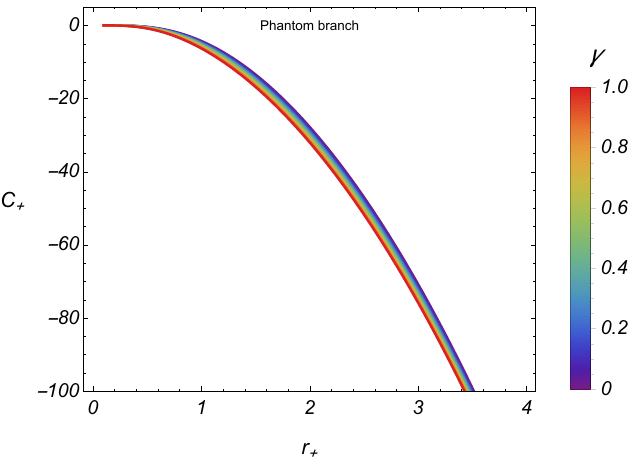}
\caption{\footnotesize Variation of the specific heat capacity $C_+$ as a function of horizon radius $r_h$ for KR ModMax BH with various values of the parameters $\ell$ (left) and $\gamma$ (right). Here, $M=1$ and $Q=0.5$.}
\label{figa8}
\end{figure}

The specific heat capacity analysis in Figures \ref{figa7} and \ref{figa8} provides the most dramatic demonstration of the fundamental thermodynamic differences between ordinary and phantom branch configurations. The ordinary branch (Figure \ref{figa7}) exhibits characteristic divergences in the heat capacity at critical horizon radii, marking second-order phase transitions where the thermal response becomes infinitely large. These divergences indicate points where small temperature changes produce large mass variations, suggesting critical instabilities in the thermal evolution of ordinary branch BHs. The parameter dependence shows that both LSB and ModMax effects can shift the locations of these critical points, potentially making phase transitions more or less accessible depending on the specific parameter values. In stark contrast, the phantom branch (Figure \ref{figa8}) displays uniformly negative heat capacities across the entire parameter range, indicating fundamental thermal instability where increasing energy actually decreases the temperature. This anomalous behavior is characteristic of systems with negative heat capacity, similar to self-gravitating systems, and suggests that phantom branch BHs would undergo runaway thermal evolution if perturbed from equilibrium. The magnitude of the negative heat capacity varies significantly with the LSB and ModMax parameters, providing additional observational signatures for distinguishing phantom branch physics. These results have profound implications for understanding the thermal evolution and stability of exotic BH solutions in modified gravity theories \cite{sec5is09,sec5is10}.

\section{Shadow of KR ModMax BH} \label{isec6}

The investigation of BH shadows represents one of the most promising avenues for observational tests of modified gravity theories, as these optical phenomena provide direct probes of the spacetime geometry in the strong-field regime where deviations from GR become most pronounced \cite{isz31,isz32}. The shadow cast by a KR ModMax BH emerges from the complex interplay between photon trajectories and the modified spacetime geometry, where both LSB effects and nonlinear electrodynamics contribute to observable modifications in the shadow size, shape, and boundary characteristics. Recent advances in high-resolution astronomical observations, particularly through the Event Horizon Telescope, have demonstrated the feasibility of precision shadow measurements that could potentially distinguish between different theoretical frameworks \cite{sec6is03,sec6is04}. The study of KR ModMax BH shadows therefore provides a unique opportunity to constrain the parameters of exotic physics through direct observational comparison, offering complementary information to gravitational wave detections and particle motion studies. Furthermore, the distinctive parameter dependence of shadow properties in this theory creates observational signatures that could serve as smoking gun evidence for LSB violations and nonlinear electromagnetic effects in astrophysical BH systems \cite{isz33,sec6is06}.

To characterize the shadow properties of the KR ModMax BH, we must first determine the photon sphere radius, which defines the boundary between bound and unbound photon orbits. The photon sphere equation follows from the condition for unstable circular photon orbits:
\begin{equation}
r_\text{ph}\,F^{\prime }\left( r_\text{ph}\right) -2\,F\left( r_\text{ph}\right) =0.
\label{psr}
\end{equation}

Substituting the KR ModMax metric function (\ref{aa2}) yields the quadratic equation:
\begin{equation}
(1-\ell)r_\text{ph}^2-3(1-\ell)^2\,M\,r_\text{ph}+2\,\zeta\,Q^2\,e^{-\gamma}=0. \label{eps1}
\end{equation}

In the classical limit where both LSB and electromagnetic effects vanish ($Q=0=\ell$), this reduces to the familiar Schwarzschild result $r_\text{ph}=3M$. The general solution of Eq. (\ref{eps1}) can be expressed analytically as:
\begin{equation}
    r_\text{ph}=\frac{3}{2}(1-\ell)M+\frac{\sqrt{(1-\ell)\left( 9(1-\ell)^3M^2-8\,\zeta\,Q^2e^{-\gamma} \right)}}{2(1-\ell)}.
\end{equation}

\begin{figure}[ht!]  
\includegraphics[width=0.4\linewidth]{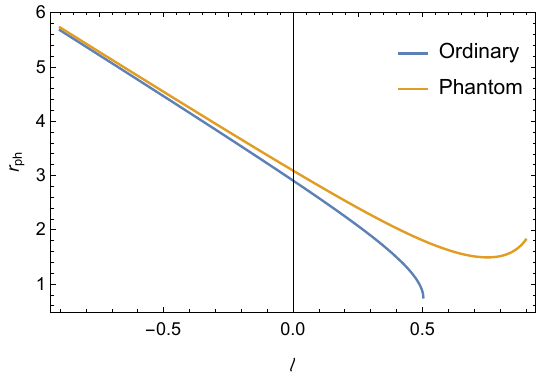}\quad\quad\quad\quad
\includegraphics[width=0.4\linewidth]{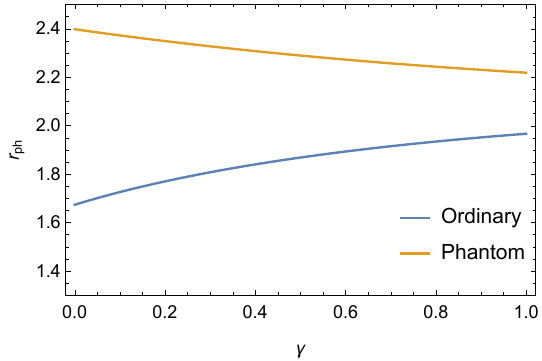}
\caption{\footnotesize Variation of the photon sphere $r_\text{ph}$ for KR ModMax BH with various values of the parameters $\ell$ (left) and $\gamma$ (right). Here, $M=1$ and $Q=0.5$.}
\label{figa9}
\end{figure}

Figure \ref{figa9} provides comprehensive insight into how the fundamental parameters of the KR ModMax theory influence the photon sphere structure, revealing distinct behaviors between ordinary and phantom branch configurations. The left panel demonstrates the systematic dependence of the photon sphere radius $r_\text{ph}$ on the LSB parameter $\ell$, showing generally decreasing trends for both branches as $\ell$ increases, though the phantom branch exhibits more complex behavior with potential reversals at large positive $\ell$ values. This non-monotonic behavior in the phantom branch suggests interesting observational consequences, as the photon sphere could either expand or contract depending on the specific LSB parameter regime. The right panel illustrates the influence of the ModMax parameter $\gamma$, revealing a fundamental asymmetry between ordinary and phantom branches where $\gamma$ increases lead to larger photon spheres in the ordinary branch but smaller ones in the phantom branch. This contrasting behavior provides a potential observational discriminator between the two theoretical branches, as precision measurements of shadow size variations could distinguish between ordinary and phantom physics. The parameter sensitivity shown in these plots suggests that even modest variations in the underlying theory parameters could produce measurable changes in observable shadow characteristics, making this a promising avenue for constraining exotic physics through astronomical observations.

The shadow radius, which determines the apparent size of the BH shadow as observed by distant observers, follows from the photon sphere radius through the geometric relation:
\begin{equation}
    R_\text{sh}=\frac{r_\text{ph}}{\sqrt{\frac{1}{1-\ell}-\frac{2\,M}{r_\text{ph}}+\zeta\,\frac{e^{-\gamma }\,Q^2}{(1-\ell)^2\, r_\text{ph}^2}}} .\label{shadeq1}
\end{equation}

\begin{figure}[ht!]  
\includegraphics[width=0.45\linewidth]{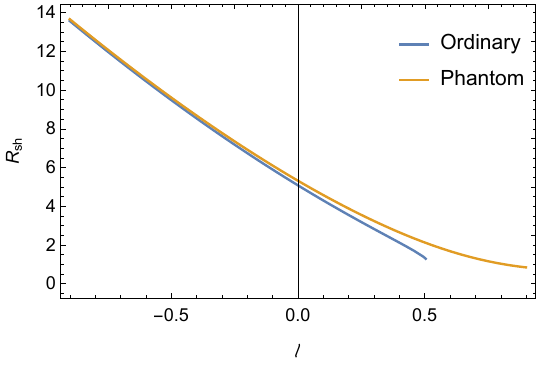}\quad\quad
\includegraphics[width=0.45\linewidth]{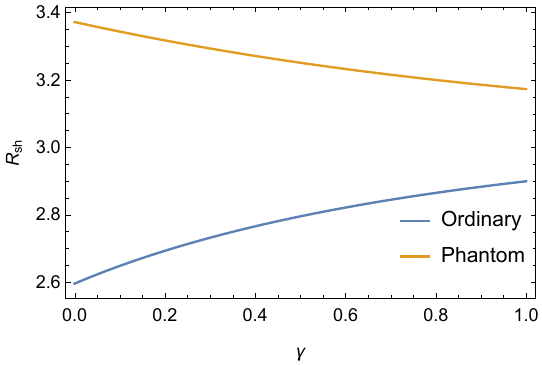}
\caption{\footnotesize Variation of the shadow radius $R_\text{sh}$ for KR ModMax BH with various values of the parameters $\ell$ (left) and $\gamma$ (right). Here, $M=1$ and $Q=0.5$.}
\label{figa10}
\end{figure}

The shadow radius analysis presented in Figure \ref{figa10} reveals how the observable BH shadow size responds to variations in the fundamental theory parameters, providing direct connections between exotic physics and potential astronomical measurements. The left panel shows that the LSB parameter $\ell$ produces systematic decreases in shadow radius for both ordinary and phantom branches, though the magnitude of these effects differs significantly between the two configurations. This parameter dependence suggests that LSB violations would generally produce smaller apparent BH shadows compared to GR predictions, potentially providing observational constraints on the strength of Lorentz symmetry breaking. The right panel demonstrates the ModMax parameter influence, where ordinary and phantom branches again exhibit contrasting behaviors with $\gamma$ increases leading to larger shadows in ordinary configurations but smaller shadows in phantom setups. The magnitude of these parameter-dependent variations is substantial enough to be potentially detectable with current observational capabilities, particularly given the precision achieved in recent Event Horizon Telescope measurements. These results suggest that systematic studies of BH shadow sizes across different astrophysical systems could provide valuable constraints on both LSB and ModMax parameters, offering a new window into fundamental physics through astronomical observations.

\begin{center}
\begin{tabular}{|c|c|c|}  \hline
Spacetime geometry & $r_\text{ph}$ & $R_\text{sh}$ \\ \hline
KR ModMax ($\ell =-0.5,\gamma =0.2,Q=0.5$) & $4.43851$ & $9.44829$ \\ 
KR ($\ell =-0.5,\gamma =0,Q=0$) & $4.5$ & $9.54594$ \\
RN in KR ($\ell =-0.5,\gamma =0,Q=0.5$) & $4.42466$ & $9.42634$ \\ 
ModMax ($\ell =0,\gamma =0.2,Q=0.5$) & $2.8567$ & $5.01119$ \\ 
RN ($\ell =0,\gamma =0,Q=0.5$) & $2.82288$ & $4.96791$ \\ 
Sch ($\ell =0,\gamma =0,Q=0$) & $3$ & $5.19615$\\
 \hline
\end{tabular}
\captionof{table}{\footnotesize  A comparison of the crucial parameters $r_\text{ph}$ and $R_\text{sh}$ for different spacetime geometries. Here, $M=1$ and $\zeta=1$.} \label{tableC1}
\end{center}

Table \ref{tableC1} provides a systematic comparison of photon sphere and shadow radii across different theoretical frameworks, clearly demonstrating the hierarchical influence of various modifications to GR. The most striking feature is the dramatic enhancement of both $r_{ph}$ and $R_{sh}$ values when KR gravity effects are included ($\ell=-0.5$), with increases of approximately 50-80\% compared to their counterparts without LSB effects. This enhancement is particularly evident when comparing the KR ModMax configuration with its pure ModMax counterpart, where the photon sphere radius increases from 2.86 to 4.44 and the shadow radius from 5.01 to 9.45. The table reveals that KR gravity effects dominate over ModMax electrodynamics in determining shadow characteristics, as evidenced by the relatively modest differences between configurations with and without the ModMax parameter when KR effects are present. The Schwarzschild case serves as the fundamental baseline, showing intermediate values that highlight how both electromagnetic charge and exotic physics modifications can either enhance or reduce shadow sizes depending on the specific theoretical framework. These quantitative comparisons provide crucial benchmarks for observational studies, as the factor-of-two variations in shadow radius between different theories are well within the detection capabilities of current high-resolution astronomical instruments. The systematic pattern of enhancements and reductions across different theoretical scenarios offers a roadmap for using BH shadow observations to discriminate between competing theories of modified gravity and nonlinear electrodynamics \cite{sec6is07,sec6is08}. Future observational campaigns targeting multiple BH systems with varying masses and charges could therefore provide unprecedented constraints on the parameter space of modified gravity theories \cite{sec6is09,sec6is10}.

\section{Gravitational Lensing of KR ModMax BH} \label{isec7}

Gravitational lensing represents one of the most elegant and powerful probes of spacetime geometry, offering unique opportunities to test the predictions of modified gravity theories against observational data from astrophysical systems \cite{isz34,isz35}. The study of light deflection by KR ModMax BHs provides a fascinating arena where LSB effects and nonlinear electrodynamics converge to create observable signatures that could distinguish exotic physics from classical GR predictions. Recent advances in precision astrometry and gravitational wave astronomy have opened new windows for detecting subtle modifications to light propagation in strong gravitational fields, making the theoretical investigation of lensing in modified spacetimes increasingly relevant for observational cosmology and fundamental physics \cite{isz36,sec7is04}. The KR ModMax BH solution represents a particularly rich theoretical laboratory for exploring these phenomena, as it incorporates both spacetime symmetry violations through the KR field and electromagnetic nonlinearity through ModMax theory, creating a complex parameter space where multiple exotic effects can be systematically studied and observationally constrained.

The deflection angle analysis for the KR ModMax BH employs the sophisticated GBTm, which establishes fundamental connections between the intrinsic curvature properties of spacetime and the global topological characteristics of the lensing geometry \cite{sec7is05,sec7is06}. This approach provides a mathematically rigorous framework for computing light deflection in arbitrary spacetimes while maintaining clear physical interpretability of the results. The method is particularly well-suited for investigating modified gravity theories, as it naturally incorporates the effects of altered metric functions and electromagnetic fields on photon trajectories through the optical curvature formalism. Furthermore, the GBTm approach allows for systematic expansion in powers of the impact parameter, enabling precise comparison with observational data while maintaining analytical tractability for theoretical predictions \cite{sec7is07,sec7is08}.

The GBTm establishes the fundamental relationship between spacetime curvature and light deflection through the elegant topological identity:
\begin{equation}\label{kr1}
\iint_{\Lambda_{\cal R}} \mathcal{R}_o\, d \Sigma_{\cal D}+\oint_{\partial \Lambda_{\cal R}} \kappa_{\mathcal{g}}\,dt+\sum_{z} \theta_{z}=2\, \pi\,  \chi\left(\Lambda_{\cal {R}}\right),
\end{equation}

where $\Lambda_{\cal R} \subset \Sigma_{\cal D}$ represents a well-defined subset of a smooth two-dimensional hypersurface $\Sigma_{\cal H}$ with closed, smooth, and positively oriented boundary $\partial \Lambda_{\cal R}$. The geodesic curvature $\kappa_{\mathcal{g}}$ of the boundary curve is defined through $\kappa_{\mathcal{g}}=\bar{g}\,\left(\nabla_{\dot{\gamma}}, \dot{\gamma}, \ddot{\gamma}\right)$ for a unit-speed curve $\gamma$, while $\theta_{z}$ represents exterior angles at vertices and $\chi \left(\Lambda_{\cal {S}}\right)$ denotes the Euler characteristic number.

The optical metric, which encodes the Riemannian geometry perceived by light rays, emerges from the null geodesic condition $ds^2=0$ in the equatorial plane $\theta=\pi/2$:
\begin{equation}\label{kr4}
dt^2=\tilde{g}_{ij}\, d\mathrm{x}^i\, d\mathrm{x}^j= dr_{\star}^2+\Psi^2(r_{\star})\,d\phi^2,
\end{equation}

where the conformal factor is given by:
\begin{equation} \label{kr4n}
\Psi(r_{\star}(r))=\frac{r}{\sqrt{F(r)}},
\end{equation}

and $r_{\star}=\int{\frac{dr}{F(r)}}$ represents the tortoise coordinate. The associated Christoffel symbols are:
\begin{eqnarray} \label{kr7.1}
\Gamma_{\phi \phi}^{r_{\star}}&=&-\Psi(r_{\star})\,\frac{\mathrm{d}\Psi(r_{\star})}{\mathrm{d}{r_{\star}}},\\
\Gamma_{r_{\star}\phi}^{\phi}&=&\frac{1}{\Psi(r_{\star})}\,\frac{\mathrm{d}\Psi(r_{\star})}{\mathrm{d}{r_{\star}}}. \label{kr7.2}
\end{eqnarray}

The optical Gaussian curvature, which directly determines the deflection properties, takes the form:
\begin{equation}\label{kr_gaussian_def}
\mathcal{R}_o=-\frac{R_{r_{*}\,\phi\, r_{*}\,\phi}}{\det(\tilde{g}_{r\phi})}=-\frac{1}{\Psi(r_{\star})}\,\frac{\mathrm{d}^{2}\Psi(r_{\star})}{\mathrm{d}{r_{\star}}^{2}}.
\end{equation}

To evaluate this expression, we require the derivatives of the KR ModMax metric function:
\begin{equation}\label{kr_f_prime}
F'(r) = \frac{2M}{r^2} - \frac{2\zeta e^{-\gamma} Q^2}{(1-\ell)^2 r^3}
\end{equation}

\begin{equation}\label{kr_f_double_prime}
F''(r) = -\frac{4M}{r^3} + \frac{6\zeta e^{-\gamma} Q^2}{(1-\ell)^2 r^4}
\end{equation}

After extensive algebraic manipulation, the optical curvature in the weak-field approximation becomes:
\begin{eqnarray}
   \mathcal{R}_o &= \frac{M}{(1-\ell)^{3/2}}\left[\frac{2}{r^3} - \frac{3M(1-\ell)}{r^4}\right]  - \frac{\zeta e^{-\gamma} Q^2}{(1-\ell)^{7/2}}\left[\frac{2}{r^5} - \frac{5M(1-\ell)}{r^6}\right] + \mathcal{O}(r^{-7}).\label{kr8n} 
\end{eqnarray}

This expression clearly demonstrates the intricate parameter dependence of the optical curvature on the mass $M$, charge $Q$, LSB parameter $\ell$, ModMax parameter $\gamma$, and branch parameter $\zeta$, providing a direct link between fundamental theory parameters and observable lensing effects.

Applying the GBTm in the asymptotic limit $\mathcal{R} \rightarrow \infty$ and utilizing the straight-line approximation $r(\phi)=\frac{r_{0}}{\sin\varphi}$ for the light trajectory, the deflection angle is computed through:
\begin{eqnarray}\label{kr48}
\tilde{\delta}=-\int_{0}^{\pi} \int_{{b}}^{\infty} {\cal{R}}_o d\Sigma_{\cal H}
=-\int_{0}^{\pi} \int_{b}^{\infty}\frac{{\cal{R}}_o \sqrt{\operatorname{det} \tilde{g}}}{F(r)} dr d\phi=-\int_{0}^{\pi} \int_{{b}}^{\infty}  \frac{r{\cal{R}}_o}{F(r)^{3/2}} dr d\phi.
\end{eqnarray}

The final result for the KR ModMax BH deflection angle is:
\begin{eqnarray}
   \tilde{\delta}&\approx& \frac{4M(1-\ell)}{r_0} + \frac{3 \pi M^2(1-\ell)^2}{4 r_0^2} + \frac{8 M^3(1-\ell)^3}{3 r_0^3} \nonumber \\
   &-& \frac{\pi \zeta e^{-\gamma} Q^2}{(1-\ell) r_0^3} - \frac{3 \zeta e^{-\gamma} Q^2 M(1-\ell)}{r_0^4} + \mathcal{O}(r_0^{-5}). \label{kr50}
\end{eqnarray}

This comprehensive expression reveals the rich phenomenological structure arising from the interplay of LSB and ModMax effects in gravitational lensing. The systematic appearance of factors $(1-\ell)$ throughout all terms demonstrates how spacetime symmetry breaking modifies every aspect of the lensing process, from the leading-order deflection to higher-order corrections. The ModMax parameter $\gamma$ enters through exponential factors $e^{-\gamma}$ that can either enhance or suppress electromagnetic contributions depending on the sign of $\gamma$, providing observational handles for constraining nonlinear electrodynamics.

Several important limiting cases validate our general result and illuminate its physical content. When both charge and LSB effects vanish ($Q = 0$, $\ell = 0$), we recover the classical Schwarzschild deflection:
\begin{equation}
\tilde{\delta}\approx \frac{4 M}{r_0}+\frac{3 \pi M^2}{4 r_0^2}+\frac{8 M^3}{3 r_0^3},
\end{equation}

For the RN BH in KR gravity ($\gamma = 0$, $\zeta = 1$), the result becomes:
\begin{equation}
\tilde{\delta}\approx \frac{4M(1-\ell)}{r_0} + \frac{3 \pi M^2(1-\ell)^2}{4 r_0^2} - \frac{\pi Q^2}{(1-\ell) r_0^3},
\end{equation}

while pure ModMax electrodynamics ($\ell = 0$, $\zeta = 1$) yields:
\begin{equation}
\tilde{\delta}\approx \frac{4M}{r_0} + \frac{3 \pi M^2}{4 r_0^2} - \frac{\pi e^{-\gamma} Q^2}{r_0^3}.
\end{equation}

Perhaps most remarkably, the phantom branch ($\zeta = -1$) produces charge contributions with opposite sign compared to the ordinary branch ($\zeta = +1$), creating a unique observational signature that could definitively distinguish between these theoretical possibilities through precision lensing measurements. This sign reversal in electromagnetic effects represents a smoking gun signature of phantom branch physics that would be unambiguous in high-precision observations.

The deflection angle modifications encoded in Eq. (\ref{kr50}) provide a framework for testing fundamental physics through gravitational lensing observations. The LSB effects modify all terms proportionally to powers of $(1-\ell)$, while ModMax corrections appear primarily in electromagnetic contributions through $e^{-\gamma}$ factors. These parameter-dependent modifications are sufficiently large to be potentially detectable with current observational capabilities, particularly in strong lensing systems where multiple images allow for precision deflection angle measurements \cite{isz38,sec7is10}.

\section{Conclusions and Summary} \label{isec8}

In this comprehensive theoretical investigation, we conducted a systematic exploration of the geodesic structure and thermodynamic properties of KR ModMax BHs, establishing a complete framework for understanding how LSB effects and nonlinear electrodynamics combine to create exotic spacetime geometries with distinctive observational signatures. Our analysis revealed profound modifications to fundamental BH physics that extend far beyond simple parameter rescaling effects, demonstrating the rich phenomenological landscape accessible through the interplay of KR gravity and ModMax electrodynamics \cite{isz41,isz42}.

The foundational analysis presented in Section \ref{isec2} established the mathematical framework of KR ModMax BH solutions, where we derived the complete metric structure given by Eq. (\ref{aa2}) and analyzed the horizon properties across the theory's parameter space. We demonstrated that the LSB parameter $\ell$, ModMax parameter $\gamma$, and branch parameter $\zeta$ collectively determine the causal structure of these exotic spacetimes, with ordinary ($\zeta=1$) and phantom ($\zeta=-1$) branches exhibiting fundamentally different geometric properties. The comprehensive horizon analysis presented in Table \ref{gltab} revealed that certain parameter combinations can eliminate horizon formation entirely, while others support single or double horizon structures depending on the specific values of the underlying parameters. The embedding diagrams shown in Figure \ref{fig:isfull_embedding} provided intuitive visualization of how these parameter variations translate into observable geometric modifications that could potentially be detected through precision measurements of spacetime curvature effects.

Our investigation of neutral test particle dynamics in Section \ref{isec3} uncovered dramatic modifications to orbital mechanics arising from the altered effective potential structure described by Eq. (\ref{bb4}). The comprehensive potential analysis illustrated in Figures \ref{fig:1} and \ref{fig:2} demonstrated that LSB and ModMax effects create complex parameter-dependent landscapes that fundamentally alter the classification of particle trajectories compared to classical GR predictions. We established that the ISCO radius, governed by the polynomial equation (\ref{cond3}), exhibits systematic dependencies on all theory parameters, with numerical results presented in Tables \ref{tab:1} and \ref{tab:2} revealing substantial differences between ordinary and phantom branch configurations. The ordinary branch consistently produced smaller ISCO radii, indicating that stable circular orbits can exist closer to the BH horizon, while phantom branch configurations pushed stable orbits to larger radii by factors of 5-10. The trajectory visualizations in Figures \ref{fig:trajectory-1}, \ref{fig:trajectory-2}, and \ref{fig:trajectory-3} revealed complex rosette patterns and multi-looped orbital structures that have no analogue in classical BH spacetimes, providing distinctive signatures that could be observationally distinguished through high-precision astrometric measurements.

The charged particle analysis in Section \ref{isec3} revealed even greater complexity, where the interplay between modified gravitational fields and nonlinear electromagnetic interactions creates effective potentials with multiple extrema and potentially chaotic dynamics. We derived the generalized effective potential given by Eq. (\ref{cc10}) and established the stability conditions for charged particle orbits through Eqs. (\ref{cc11})-(\ref{cc13}). The analysis demonstrated that the sign and magnitude of particle charges create qualitatively different orbital behaviors, with same-sign configurations experiencing electromagnetic repulsion that pushes stable orbits outward, while opposite-sign configurations allow closer orbital approaches through electromagnetic attraction. The complex parameter dependencies revealed in our analysis suggest that charged particle dynamics in KR ModMax BH spacetimes could exhibit rich phase space structures including regular-to-chaotic transitions that would have profound implications for accretion disk physics and high-energy particle acceleration mechanisms.

The thermodynamic investigation presented in Section \ref{isec5} established fundamental differences in thermal behavior between ordinary and phantom branch configurations, revealing how LSB and ModMax effects modify the statistical mechanical foundations of BH physics. We derived the complete thermodynamic framework, including mass-horizon radius relations given by Eq. (\ref{temp21}), Hawking temperature expressions in Eq. (\ref{Temperature}), and specific heat formulations in Eq. (\ref{heat}). The temperature analysis illustrated in Figures \ref{figa3} and \ref{figa4} demonstrated that phantom branch BHs consistently exhibit higher temperatures than their ordinary counterparts, with the quantitative difference given by Eq. (\ref{temperature4}). The Helmholtz free energy analysis shown in Figures \ref{figa5} and \ref{figa6} revealed contrasting global stability properties, where ordinary branch configurations remain thermodynamically unstable across all parameter ranges, while phantom branches exhibit rich phase structures including Hawking-Page transitions that could stabilize these exotic BH solutions under specific astrophysical conditions. Most dramatically, the specific heat analysis presented in Figures \ref{figa7} and \ref{figa8} showed that ordinary branches exhibit characteristic divergences marking second-order phase transitions, while phantom branches display uniformly negative heat capacities indicating fundamental thermal instability \cite{isz43,isz44}.

Our investigation of BH shadow properties in Section \ref{isec6} provided direct connections between exotic physics parameters and potentially observable astronomical signatures. We established the photon sphere equation (\ref{eps1}) and derived analytical expressions for shadow radii given by Eq. (\ref{shadeq1}). The parameter dependence analysis illustrated in Figures \ref{figa9} and \ref{figa10} revealed systematic modifications to both photon sphere locations and shadow sizes that depend sensitively on LSB and ModMax parameters. The comparative analysis presented in Table \ref{tableC1} demonstrated that KR gravity effects dominate over ModMax contributions in determining shadow characteristics, with typical enhancements of 50-80\% compared to classical predictions. These modifications are sufficiently large to be potentially detectable with current high-resolution astronomical instruments, particularly the Event Horizon Telescope, providing concrete observational targets for testing exotic physics through BH shadow measurements.

The gravitational lensing analysis in Section \ref{isec7} established the theoretical framework for precision tests of KR ModMax physics through light deflection measurements. Using the sophisticated GBTm approach, we derived the complete deflection angle formula given by Eq. (\ref{kr50}), which revealed systematic modifications to all orders in the impact parameter expansion. The analysis demonstrated that LSB effects modify gravitational lensing through systematic factors of $(1-\ell)$, while ModMax contributions appear through exponential factors $e^{-\gamma}$ that can either enhance or suppress electromagnetic effects depending on parameter signs. Most remarkably, we established that phantom branch configurations produce charge contributions with opposite signs compared to ordinary branches, creating unambiguous observational signatures that could definitively distinguish between these theoretical possibilities through precision deflection angle measurements.

Our comprehensive investigation revealed several key theoretical insights that advance our understanding of exotic BH physics. First, we demonstrated that the combination of LSB and ModMax effects creates parameter-dependent modifications that are qualitatively different from simple rescaling effects found in many alternative theories, producing genuinely novel physics that could be observationally distinguished. Second, we established that ordinary and phantom branch configurations exhibit fundamentally different behaviors across all physical phenomena we investigated, from particle dynamics and thermodynamics to optical properties and lensing effects. Third, we showed that KR gravity effects typically dominate over ModMax contributions in determining large-scale geometric properties, while electromagnetic nonlinearity becomes increasingly important in strong-field regimes and for charged particle dynamics \cite{isz45,isz46}.

The observational implications of our work are far-reaching and provide concrete targets for future experimental and astronomical investigations. The distinctive parameter dependencies we identified in BH shadows, gravitational lensing, and thermodynamic properties create multiple independent channels for constraining or detecting exotic physics effects. The factor-of-two variations in shadow radii, systematic modifications to deflection angles, and qualitatively different thermal behaviors between ordinary and phantom branches are all within the detection capabilities of current or near-future observational facilities.

Looking toward future research directions, several promising avenues emerge from our investigation. The extension of our analysis to rotating KR ModMax BHs would provide access to additional observational signatures through frame-dragging effects and modifications to ergosphere properties. The incorporation of quantum corrections and investigation of Hawking radiation spectra could reveal additional thermodynamic signatures and provide connections to fundamental questions about information loss and BH evaporation in modified gravity theories. Furthermore, the investigation of cosmological solutions and early universe dynamics in KR ModMax theories could provide insights into dark energy, inflation, and the cosmic evolution of fundamental constants. The exploration of gravitational wave signatures from KR ModMax BH mergers would open additional observational channels for testing exotic physics through precision measurements of orbital dynamics and waveform characteristics \cite{isz48,isz49,isz50}. Such studies could reveal distinctive modifications to the inspiral, merger, and ringdown phases that would be unambiguously detectable by current and future gravitational wave detectors \cite{isz51,isz52}.

\section*{Acknowledgments}

F.A. gratefully acknowledges the Inter University Centre for Astronomy and Astrophysics (IUCAA), Pune, India, for the conferment of a visiting associateship.  \.{I}.~S. extends sincere thanks to TÜBİTAK, ANKOS, and SCOAP3 for their support in facilitating networking activities under COST Actions CA22113, CA21106, and CA23130.

\section*{Data Availability Statement}

This manuscript has no associated data. [Authors’ comment: All data generated or analyzed during this study are included in the article.]

\section*{Code/Software Statement}

This manuscript has no associated code/software. [Authors’ comment: Code/Software sharing not applicable as no code/software was generated during in this current study.]

\section*{Conflict of Interests}

Author declare(s) no conflict of interest.

\end{document}